# Masgent: An AI-assisted Materials Simulation Agent

Guangchen Liu[1], Songge Yang[1], Yu Zhong[1,*]

[1]Mechanical and Materials Engineering Department, Worcester Polytechnic Institute, 100 Institute Rd, Worcester, MA, 01609, USA

[*]Corresponding author: Yu Zhong; E-mail: yzhong@wpi.edu

## Abstract

*Density functional theory (DFT) and machine learning potentials (MLPs) are essential for predicting and understanding materials properties, yet preparing, executing, and analyzing these simulations typically requires extensive scripting, multi-step procedures, and significant high-performance computing (HPC) expertise. These challenges hinder reproducibility and slow down discovery. Here, we introduce* ***Masgent****, an AI-assisted materials simulation agent that unifies structure manipulation, automated VASP input generation, DFT workflow construction and analysis, fast MLP-based simulations, and lightweight machine learning (ML) utilities within a single platform. Powered by large language models (LLMs), Masgent enables researchers to perform complex simulation tasks through natural-language interaction, eliminating most manual scripting and reducing setup time from hours to seconds. By standardizing protocols and integrating advanced simulation and data-driven tools, Masgent lowers the barrier to performing state-of-the-art computational methodologies, enabling faster hypothesis testing, pre-screening, and exploratory research for both new and experienced practitioners.*

## Program Summary

**Program Title:** Masgent

**Developer:** Guangchen Liu

**Repository:** https://github.com/IMPDGroup/Masgent

**PyPI Distribution:** https://pypi.org/project/Masgent

**License:** MIT

**Programming Language:** Python

**Nature of Problem:** Density functional theory (DFT) workflows, machine learning potential (MLP) simulations, and data-driven materials modeling typically require extensive scripting, multi-step procedures, and significant high-performance computing (HPC) expertise. These complexities make advanced simulations difficult to set up, reproduce, and analyze, particularly for new users.

**Solution Method:** Masgent is an AI-assisted framework that automates structure manipulation, VASP input generation, DFT workflow construction and analysis, fast MLP simulations, and lightweight machine learning (ML) tasks for materials science. Through natural-language interaction powered by large language models (LLMs), Masgent reduces setup time from hours to seconds, standardizes protocols, and improves reproducibility, lowering technical barriers and accelerating computational materials research.

## 1. Introduction

Density functional theory (DFT) simulations and machine learning potentials (MLPs) have become indispensable tools in materials science for predicting and understanding materials properties, screening candidate compounds, and guiding experimental design [1-9]. Despite their widespread adoption, the practical implementation of these simulations remains challenging. Even standard workflows, such as structural relaxations, equation of state (EOS) evaluations, or elastic property calculations, require extensive scripting, careful input preparation, detailed knowledge of simulation software (e.g., VASP [10, 11]), and expertise in high-performance computing (HPC) environments. These complexities create barriers for new users, complicate reproducibility, and hinder the rapid prototyping and iterative exploration essential for modern materials research.

A broad ecosystem of tools has emerged to simplify portions of these workflows. Libraries like Pymatgen [12] and ASE [13] provide powerful APIs for structure manipulation and input generation, while tools like VASPkit [14], qvasp [15], and Atomate [16, 17] support pre-processing, workflow construction, and post-processing of VASP calculations. Machine learning frameworks, including Scikit-learn [18], TensorFlow [19], and PyTorch [20], enable data-driven modeling and property prediction, and large-scale materials databases such as the Materials Project [21] and Open Quantum Materials Database (OQMD) [22, 23] provide curated structural and thermodynamic data. Recent advances in MLPs, such as SevenNet [24], CHGNet [25], Orb-v3 [26, 27], MatterSim [28], M3GNet [29], MACE [30], and Matlantis [31], have further accelerated materials simulations by providing near-DFT accuracy at orders-of-magnitude lower cost. However, these tools remain largely fragmented, require significant manual integration, and often demand significant technical expertise to deploy effectively. As summarized in **Table 1**, existing tools remain fragmented, none provide a unified environment that seamlessly integrates DFT automation, MLP simulations, and machine learning (ML) modeling within an intuitive, user-friendly interface. At the same time, the rise of artificial intelligence (AI) and agentic systems [32-38] has highlighted the potential for natural-language interfaces that allow researchers to interact with simulation tools more intuitively and efficiently.

To address these challenges, we introduce **Masgent**, an AI-assisted materials simulation agent designed to lower the barrier to performing advanced computational techniques. Masgent unifies

DFT automation, fast MLP simulations, and lightweight ML utilities into a single, extensible framework. At its core is an interactive AI agent that enables natural-language interaction, allowing users to prepare simulations, generate inputs, and construct workflows without extensive scripting. The framework automates key tasks such as structure manipulation, defect generation, supercell construction, surface and interface creation, and complete VASP input preparation for different scenarios such as relaxation and static calculations. It further provides standardized workflows for common simulations, such as convergence tests, EOS fitting, elasticity evaluations, ab initio molecular dynamics (AIMD) simulations, and nudged elastic band (NEB) calculations, along with built-in analysis tools for extracting relevant properties. Masgent additionally incorporates high-performance MLP engines, including SevenNet, CHGNet, Orb-v3, and MatterSim, enabling rapid estimation of material properties at a fraction of the computational cost of traditional DFT. These capabilities enable near-instantaneous testing of hypotheses, pre-screening of candidate structures, and exploratory simulations that would otherwise be prohibitively expensive. Beyond physical-based simulations, Masgent includes lightweight ML utilities for data preparation, feature analysis, dimensionality reduction, data augmentation, hyperparameter tuning, model training and evaluation, and pre-trained model applications, bridging traditional simulations and emerging data-driven materials research.

By integrating these diverse functionalities into a coherent platform, Masgent significantly lowers the barrier to performing advanced materials simulations. Users can prepare, modify, and analyze computational workflows using conversational commands, eliminating much of the boilerplate scripting and manual file manipulation traditionally required. This design improves accessibility for students and newcomers while accelerating productivity for expert practitioners. The framework additionally promotes reproducibility through standardized input templates, validated tool calls, and structured workflow organization. In the following sections, the architecture and capabilities of Masgent are described in detail, demonstrating its utility through representative case studies, and discussing its broader impact on materials simulation and autonomous scientific computing.

## 2. Architecture and Design

Masgent is designed as a modular, extensible, and AI-native framework that unifies DFT automation, MLP simulations, ML utilities, and natural-language interactions into a cohesive software ecosystem. The architecture follows a layered design that cleanly separates user interaction from computational backends, enabling flexible use across conversational sessions, scripted workflows, and high-throughput pipelines. The overall structure is illustrated in **Figure 1**.

At the highest level, Masgent consists of three interconnected components:

(1) An AI-driven natural-language interface

(2) A deterministic, menu-based command-line interface (CLI)

(3) A core utilities layer implementing all computational functionality

This separation ensures both usability and reproducibility: users may interact with the system through natural language, while all underlying operations remain structured, validated, and fully deterministic.

**2.1 AI Mode (src/ai_mode):**

The AI mode provides Masgent's natural-language interface, enabling users to request simulation tasks conversationally. The LLM interprets user instructions and maps them to specific tools through a structured tool-calling system. The LLM is guided by robust system prompts that regulate its behavior, ensuring reliability and preventing unsafe or ambiguous actions.

Each tool is associated with a Pydantic schema, which precisely defines required inputs, permissible values, and strict type constraints. This guarantees that all operations invoked through AI mode undergo rigorous validation before execution. The AI agent also maintains conversational context, allowing users to refine requests over multiple turns, ask follow-up questions, and receive intelligent feedback on planned operations or missing information.

**2.2 CLI Mode (src/cli_mode):**

The CLI mode provides an alternative deterministic interface built around a menu-driven workflow. Using the Bullet library, users can navigate a structured set of commands to perform common tasks without writing scripts or interacting with the AI agent. This mode is particularly valuable for users who prefer a guided experience or when operating in environments where natural-language processing may be impractical.

The CLI uses the same core utilities as AI mode, ensuring that all interaction pathways produce consistent and reproducible outcomes. When Masgent is launched, the CLI serves as the default entry point, and the AI mode may be activated at any point during a session whenever more flexible or conversational interaction is desired.

**2.3 Core Utilities (src/utils):**

The Core utilities layer implements all substantive computations within Masgent and is designed to be modular, framework-agnostic, and callable from both the AI agent and the CLI. This layer includes:

- Structure manipulation tools, including vacancy and substitution defects, interstitial generation, supercell construction, Special Quasirandom Structure (SQS) generation, slab creation, and interface construction via lattice matching.

- Automated VASP input preparation, including INCAR, KPOINTS, POTCAR, and HPC job scripts based on standardized, best-practice templates.

- Workflow generators for structure relaxation, static calculations, EOS fitting, elasticity evaluation, AIMD simulations, and NEB calculations.

- Fast-simulation capabilities via unified interfaces to MLP engines such as SevenNet, CHGNet, Orb-v3, and MatterSim, enabling rapid energy evaluations, EOS, elasticity, and molecular dynamics (MD) simulations.

- Lightweight ML utilities, including feature analysis via principal component analysis (PCA) [39], data augmentation via conditional variational autoencoders (CVAEs) [40-42], hyperparameter optimization via Optuna [43], model training and evaluation, and pre-trained model applications.

Together, these core utilities encapsulate all algorithmic logic, allowing the higher-level interfaces (AI mode and CLI mode) to remain thin, maintainable, and focused on user interaction rather than computation.

## 3. Features and Capabilities

### 3.1 AI Agent

Masgent incorporates an AI-native interface that allows users to perform complex material simulation tasks through natural-language interactions. The AI agent acts as a high-level orchestrator: it interprets user intents, translates natural-language instructions into well-defined computational operations, and invokes the appropriate tools within the Masgent framework to execute these tasks. This design removes the need for extensive scripting or deep HPC expertise, enabling researchers to focus on conceptual and scientific questions rather than technical implementation details.

As shown in **Figure 2**, the agent is built on a structured tool-calling architecture implemented with Pydantic AI. Incoming user messages are processed by the selected LLM backend (**Table 2**), which may be powered by OpenAI, Anthropic, Google, DeepSeek, or other supported providers. The LLM backend identifies the intended operation, such as structure generation, defect creation, VASP input preparation, workflow construction, MLP evaluation, or results analysis, and maps it to a specific tool. The agent supports multi-step workflows by breaking down complex user requests into sequential tasks, managing dependencies, and ensuring that intermediate results are appropriately passed between steps.

Each tool in Masgent is defined by a dedicated Pydantic schema that precisely specifies valid argument types, constraints, and default behavior. These schemas enforce robust input validation, detect missing or ambiguous parameters, and ensure that all operations are syntactically and physically consistent (e.g., valid chemical formulas, sensible k-point densities, admissible supercell sizes). When ambiguity is detected, the agent explicitly asks for clarification, ensuring transparency and preventing inappropriate or unsafe execution.

The AI agent is also designed for multi-turn, context-aware interaction. It maintains lightweight session memory that tracks active structures, working directories, previous outputs, and user preferences. This allows users to reference prior results implicitly, for example: "use the relaxed structure from earlier" or "repeat that calculation with a denser k-point grid. To minimize latency and token consumption, the agent retains only contextual information required to execute ongoing tasks, while the full conversation history is still stored for each session and can be accessed or reviewed later.

The AI agent supports a broad range of capabilities, including but not limited to structure manipulation, VASP input preparation, construction of DFT workflows, fast MLP simulations, and various ML tasks. A complete list of current functionalities is provided in **Appendix A**. All interactions, tool calls, and workflow decisions are automatically logged in a structured format to ensure reproducibility and traceability. Sessions can be saved, reviewed, or shared with collaborators, facilitating transparency, knowledge transfer, and consistent computational practices across users and research groups.

To illustrate how Masgent integrates natural-language interaction with structured workflow automation, **Figure 3** presents a complete end-to-end example of the AI agent generating a full simulation workflow from a single conversational instruction. In this demonstration, a user requests a simulation workflow for $LaCoO_3$ involving three sequential tasks: (1) retrieve a pristine structure from the Materials Project; (2) introduce an oxygen vacancy; and (3) estimate the resulting defect energy using an MLP model.

As shown in **Figure 3**, the left panel displays a representative interaction between the user and the agent. The agent interprets the request, constructs a multi-step execution plan, resolves ambiguous details (e.g., defect type, number of vacancies, choice of MLP engine), and requests confirmation before proceeding. This transparent, iterative planning process ensures that all parameters are fully specified, physically meaningful, and aligned with user intent before execution.

The right panel shows the automatically generated workflow directory produced by Masgent. All intermediate and final outputs, including pristine and defective structures, MLP input and output files, and conversation transcripts, are organized into a standardized directory layout. As indicated by the timestamps, the full interaction required only about three minutes, including the user's initial request, the agent's intent parsing and planning, user confirmation, and the agent's execution and summarization of results.

This example demonstrates how Masgent transforms a high-level instruction, "Prepare a $LaCoO_3$ structure with an oxygen vacancy and estimate its energy using MLP", into a fully configured, reproducible simulation pipeline within seconds to minutes. It highlights the practical utility of AI-assisted workflow construction and its potential to significantly accelerate early-stage computational exploration.

It is important to note that while the AI agent simplifies the setup, configuration, and post-processing of simulations, it does not execute VASP calculations directly. Instead, it orchestrates the workflow surrounding these calculations. Certain expert-level tasks, such as diagnosing subtle VASP error messages or optimizing HPC job scheduling, may still require manual intervention and domain expertise. Nevertheless, by automating routine operations and providing an intuitive natural-language interface, Masgent significantly lowers the barrier to entry and enables a broader range of users to leverage advanced computational simulation techniques.

## 3.2 Structure Manipulation and Input Generation

Masgent provides a comprehensive suite of tools for structure manipulation and automated input generation for DFT simulations. These capabilities form the foundation of streamlined workflow construction, allowing users to perform complex structural operations and prepare high-quality simulation inputs with minimal manual effort. All structure-related functionalities are built upon robust foundations provided by ASE, Pymatgen, and Masgent's own utility modules, ensuring compatibility with widely used computational materials workflows.

### 3.2.1 Structure Preparation, Conversion, and Visualization

Masgent supports all major structure file formats used in atomic simulations, including VASP POSCAR/CONTCAR, CIF, and XYZ, and enables conversion between them. Structures may be loaded from local files or retrieved directly from the Materials Project via a user-provided API key, giving researchers rapid access to experimentally validated and computationally optimized crystal prototypes.

For VASP workflows, Masgent offers convenient tools to switch between Direct and Cartesian coordinates, facilitating compatibility with downstream simulation codes and analysis tools. As illustrated in **Figure 4**, Masgent also provides interactive, browser-based 3D visualization using 3Dmol.js [44], allowing users to inspect structural geometries, identify potential issues, and verify structural configurations before launching simulations. The viewer supports full rotation and zooming, and uses default atomic radii derived from VESTA [45, 46] (**Table 3**) to ensure physically meaningful and visually consistent representations of atomic structures.

### 3.2.2 Structure Construction and Modification

Masgent provides first-class support for constructing and modifying atomistic structures, enabling users to generate a wide range of configurations commonly required in materials simulations.

***i. Defect Generation:***

As shown in **Figure 5**, Masgent automates the creation of vacancy, substitutional, and interstitial defects. Vacancy creation allows users to specify either the number of atoms to remove or a target concentration, after which Masgent randomly selects atomic sites for removal. Substitutional defects are generated by randomly replacing host atoms with dopant species according to user-specified compositions. For interstitials, Masgent employs the *VoronoiInterstitialGenerator* from pymatgen-analysis-defects [47], which identifies plausible interstitial positions using Voronoi analysis and places the desired species accordingly. Together, these tools allow users to rapidly prototype defect chemistries without manually editing structure files.

***ii. Supercell Generation:***

**Figure 6a** illustrates a 2×2×2 $LaCoO_3$ supercell generated using Masgent's supercell module. Users provide an integer 3×3 scaling matrix, and Masgent automatically replicates lattice vectors and atomic positions while preserving structural symmetry. Such supercells are essential for modeling dilute defects, finite-size effects, lattice dynamics, and long-range structural distortions.

***iii. SQS Generation:***

To simulate chemically disordered systems, Masgent integrates with the Icet package [48] to generate SQS structures. As shown in **Figure 6b**, users specify the target composition, supercell size, and correlation functions to match. Icet's Monte Carlo optimization framework then produces an atomic arrangement whose statistics approximate those of a perfectly random alloy, ideal for modeling high-entropy oxides, doped perovskites, and solid solutions.

***iv. Surface Slab Generation:***

Masgent also supports automated surface slab generation from bulk structures. **Figure 7a** and **b** show examples of $LaCoO_3$ and $La_2NiO_4$ (001) slabs generated through the framework. Users specify Miller indices, slab thickness, and vacuum spacing, and Masgent constructs a periodic slab geometry with user-defined terminations. These slabs are suitable for surface energy calculations, adsorption studies, and catalysis simulations.

***v. Interface Structure Generation:***

For heterostructure and multilayer simulations, Masgent incorporates a lattice-matching algorithm [49] to construct interfaces between different materials. As demonstrated in **Figure 7c**, users specify the two bulk crystals, orientations, and matching tolerances. Masgent identifies compatible surface-pair combinations and generates a strain-minimized interface supercell. This functionality is particularly useful for studying epitaxial growth, grain boundaries, oxide heterostructures, and interfacial transport.

### 3.2.3 Automatic VASP Input Generation

Masgent provides turnkey utilities for generating high-quality VASP input files that follow best practices and established parameter standards. Users may rely on built-in templates for common simulation types or customize settings to meet specific workflow requirements. These tools significantly reduce manual setup effort while improving reproducibility and consistency across calculations.

***i. INCAR Generation:***

Masgent includes predefined INCAR templates adapted from Pymatgen and optimized for various calculation types, including *MPMetalRelaxSet* (metallic relaxations), *MPRelaxSet* (general structure relaxations), *MPStaticSet* (static energy calculations), *MPNonSCFBandSet* (non-self-consistent band structure), *MPNonSCFDOSSet* (non-self-consistent density-of-states), and *MPMDSet* (ab initio molecular dynamics). Each template provides sensible defaults derived from established Materials Project workflows while allowing advanced users to override parameters for finer control.

***ii. KPOINTS Generation:***

The KPOINTS generator creates either Monkhorst-Pack or Γ-centered meshes depending on user preference and lattice symmetry. Users can specify target accuracy levels (low, medium, or high) or provide explicit k-point densities. Masgent then automatically constructs an appropriate grid that balances computational cost and convergence quality, ensuring adequate Brillouin-zone sampling across diverse system types.

***iii. POTCAR Generation:***

Pseudopotentials are assembled from locally stored PAW datasets. Masgent automatically selects the correct POTCAR files based on the chemical elements present in the structure, ensuring compatibility and consistency with VASP's recommended datasets and preventing manual errors. This greatly simplifies setup for multi-element compounds and large defect or interface supercells.

***iv. HPC Job Script Generation:***

For HPC execution, Masgent generates ready-to-submit job scripts compatible with schedulers such as SLURM. Users can specify computational resources, including node count, tasks per node, wall time, queue/partition, and module environments, and Masgent automatically constructs a script that loads the appropriate modules and launches VASP with correct MPI settings. These scripts follow best-practice conventions for parallel performance, traceability, and job reproducibility.

### 3.3 DFT Workflow Automation and Analysis

Masgent provides a unified framework for constructing, organizing, and analyzing standard DFT workflows. These workflows encapsulate widely adopted computational protocols, including structure relaxations, static energy calculations, elasticity calculations, EOS fitting, AIMD simulations, and NEB pathways. By integrating the automated input-generation tools described in Section 3.2 with modular workflow templates and dedicated post-processing routines, Masgent enables users to conduct reproducible, systematic, and efficient DFT studies while minimizing manual intervention and reducing opportunities for human errors.

As illustrated in **Figure 8**, Masgent automatically generates a complete VASP workflow directory with clearly labeled subfolders, template input files, structured output locations, and ready-to-submit HPC job scripts. This standardized directory architecture ensures immediate usability on both local workstations and large HPC clusters.

Masgent currently provides predefined workflow templates for the following classes of DFT simulations:

***i. Convergence Tests:***

Convergence tests for key numerical parameters, such as the plane-wave energy cutoff (ENCUT) and k-point density, are critical for ensuring accuracy and reproducibility. Masgent automates these tests by generating a series of calculations with systematically varied ENCUT values or k-point meshes, preparing all required VASP inputs, and organizing the runs into structured directories. After execution, Masgent collects and analyzes energy differences per atom to identify convergence thresholds and recommend optimal simulation settings.

***ii. EOS Calculations:***

Masgent provides automated tools to compute EOS curves by generating a sequence of volume-scaled structures around the equilibrium configuration. For each scaled structure, the framework prepares static calculations and organizes them into a consistent workflow directory. Once energies are collected, Masgent fits the resulting energy-volume data to standard EOS models, such as the Birch-Murnaghan equation, to extract equilibrium volume and generate the corresponding optimized structure.

***iii. Elasticity Calculations:***

To evaluate elastic properties, Masgent constructs a set of symmetry-preserving strained structures by applying small deformations to the equilibrium lattice. Each strain state is placed in a separate directory. After the static calculations are completed, the framework extracts stress tensors and constructs the elastic stiffness matrix using linear elasticity theory.

***iv. AIMD Simulations:***

Masgent automates the preparation and organization of AIMD simulations by generating input decks for user-specified temperatures, time steps, and simulation durations. AIMD workflows are organized by temperature. Post-processing utilities compute trajectory statistics such as temperature equilibration, energy fluctuations, mean-squared displacement (MSD), diffusion coefficients, and Arrhenius behavior.

***v. NEB Calculations:***

For migration-barrier and reaction-pathway studies, Masgent automates the NEB workflow by generating intermediate images between initial and final states using interpolation algorithms. Each image is placed into a numbered directory. The framework produces complete job scripts and input templates suitable for climbing-image NEB (CI-NEB) or standard NEB calculations. After execution, Masgent provides tools to extract reaction-coordinate positions, compute image energies, construct the full energy profile, and evaluate the corresponding migration barrier.

### 3.4 Fast Machine Learning Potential Simulations

Masgent integrates several state-of-the-art MLPs to enable rapid, low-cost simulations that approximate DFT-level accuracy. These capabilities dramatically accelerate preliminary screening, large-scale simulations, and workflow prototyping, often reducing compute time by orders of magnitude relative to conventional DFT. As illustrated in **Figure 9**, Masgent provides a unified MLP framework that abstracts away the differences between individual models, allowing users to switch seamlessly between potentials while preserving a consistent workflow interface.

***i. Unified MLP Interface:***

Masgent's MLP module wraps multiple widely used models, including SevenNet, CHGNet, Orb-v3, and MatterSim, under a single, standardized API. Although these models differ in architecture, training data, and target material domains, Masgent exposes a uniform set of high-level function calls for energy evaluation, EOS fitting, elasticity calculations, and MD simulations. Users simply select the desired MLP engine, and the remaining workflow logic remains unchanged. This unified interface simplifies experimentation and enables cross-model benchmarking without modifying code or directory structures.

***ii. Single-Point and EOS Calculations:***

For rapid energetic assessments, Masgent supports single-point energy and force evaluations using any supported MLP engines. For EOS calculations, Masgent automatically generates volume-scaled structures, evaluates their energies with the selected potential, and fits the resulting energy-volume curve to standard EOS models. This yields equilibrium volumes and optimized structures at a fraction of the cost of DFT, enabling quick exploration of structural stability and material trends before committing to more computationally intensive first-principles calculations.

***iii. Elasticity Calculation:***

Masgent provides MLP-accelerated elasticity workflows by applying small strains to a structure, computing the resulting stress tensors via the chosen potential, and extracting elastic constants using linear elasticity theory. These calculations typically complete in seconds to minutes, allowing high-throughput screening of mechanical properties across large compositional or configurational design spaces.

***iv. MD Simulations:***

Using the same unified interface, Masgent enables MLP-driven MD simulations in the NVT ensemble using a Nosé-Hoover chain thermostat. Users may specify temperature, timestep, and total simulation duration. The framework handles trajectory generation and provides post-processing tools to compute MSD, diffusion coefficients, and temperature and energy evolution. These MLP-based MD simulations offer orders of magnitude faster than AIMD, making them particularly useful for exploratory phase-space sampling, transport property analysis, and preliminary stability assessments.

### 3.5 Lightweight Machine Learning Utilities

In addition to automated DFT and MLP workflows, Masgent provides a suite of lightweight ML utilities designed to support materials informatics tasks such as from-scratch model development and pre-trained model applications. As shown in **Figure 10**, these tools enable researchers and students to rapidly prototype ML workflows directly within Masgent, eliminating the need for external environments and offering a unified platform that integrates both physics-based and data-driven methodologies.

***i. Feature Exploration and Correlation Analysis:***

Masgent provides utilities for loading structured datasets (e.g., CSV files containing descriptors and target properties) and performing exploratory feature analysis. Users can compute a correlation matrix heatmap to visualize linear relationships, redundant descriptors, and potential descriptor groups relevant for model construction. All plots are produced in publication-quality formats and may be exported for further use in manuscripts, presentations, or reports.

***ii. Dimensionality Reduction:***

To help interpret high-dimensional datasets, common in composition-property mappings, defect thermodynamics studies, and doping analyses, Masgent incorporates a built-in PCA module. PCA projects data into a lower dimension while preserving dominant variance directions, allowing users to visualize clustering behavior, flag outliers, and identify low-dimensional manifolds that may inform subsequent modeling choices.

***iii. Data Augmentation:***

For scenarios where training data are sparse, a frequent challenge in materials science research, Masgent provides a lightweight CVAE implementation. The CVAE module generates synthetic samples conditioned on existing descriptors, producing augmented datasets that more accurately reflect the distribution of the training set. This helps mitigate overfitting and improve the robustness of downstream predictive models.

***iv. Hyperparameter Tuning:***

Masgent integrates Optuna-based hyperparameter optimization to streamline model selection. Users define the hyperparameter search spaces for a given ML model, and Masgent employs Bayesian optimization, via the Tree-Structured Parzen Estimator (TPE) [50], to efficiently identify the best-performing configurations. This automation automates the search procedure, reduces manual trial-and-error, and improves overall predictive performance across diverse modeling tasks.

***v. Model Training and Evaluation:***

Masgent provides simplified interfaces for training, validating, and evaluating ML models. Standardized routines support train-test splitting, early stopping with patience, loss and accuracy tracking, and evaluation metrics such as RMSE and $R^2$. The framework also generates training-curve visualizations that help diagnose model convergence and generalization behavior. These tools allow users to construct baseline predictive models for properties such as formation energies, lattice distortions, diffusion barriers, and defect energetics with minimal setup and without requiring advanced ML expertise.

***vi. Pre-Trained Model Applications***

In addition to from-scratch model development, Masgent supports direct property prediction using pre-trained ML models. These models are adapted from established studies on Sc-modified Al-Mg-Si alloys [51, 52] and Al-Co-Cr-Fe-Ni high-entropy alloys (HEAs) [53], and are embedded within the framework. For Sc-modified Al-Mg-Si alloys, users provide elemental concentrations (e.g., Mg and Si), and Masgent returns CALPHAD-derived phase fractions of the primary Al phase, eutectic phases, and the $AlSc_2Si_2$ phase, as well as predicted mechanical properties such as ultimate tensile strength (UTS), yield strength (YS), and elongation (EL). For Al-Co-Cr-Fe-Ni HEAs, users input elemental concentrations (e.g., Al, Co, Cr, and Fe), and Masgent predicts phase stability metrics such as formation enthalpy, along with elastic properties including elastic constants and elastic moduli. By combining pre-trained inference with from-scratch training utilities, Masgent provides a flexible ML ecosystem that enables both rapid property screening and customized model development within a unified environment.

## 4. Benchmarks and Case Studies

To evaluate the reliability, efficiency, and practical utility of Masgent, a series of benchmarks and case studies were performed, covering structure preparation, automated DFT workflow generation, MLP simulations, and lightweight ML tasks. Detailed input files, raw results, and additional examples are available at https://github.com/IMPDGroup/Masgent/tree/main/examples. These demonstrations highlight Masgent's ability to accelerate simulation setup, reduce human error, and provide rapid scientific insight across diverse material systems.

### 4.1 Workflow Preparation Speed and Efficiency

We benchmarked Masgent's workflow preparation speed by measuring the time required to set up standard DFT simulations for a representative set of materials. Using the natural-language AI agent, complete simulation workflows, including structure retrieval, manipulation, input file creation, and directory organization, were automatically generated for ten diverse systems: simple metals (Al, Cu), elemental and compound semiconductors (Si, GaAs, $MoS_2$), ionic and transition-metal oxides (MgO, NiO, $TiO_2$), and complex oxide frameworks such as perovskites and Ruddlesden-Popper phases ($LaCoO_3$, $La_2NiO_4$).

Masgent prepared full workflows for each system in under 30 seconds, representing a dramatic improvement compared to manual scripting approaches, which typically require 1-3 hours per material depending on workflow complexity. The AI agent's ability to interpret user intent, construct multi-step plans, and automatically resolve missing parameters not only accelerates productivity but also reduces cognitive burden and minimizes the likelihood of human-introduced errors during simulation setup.

### 4.2 MLPs Performance Benchmarks

To assess the accuracy and computational efficiency of Masgent's fast-simulation module, we benchmarked the supported MLPs, SevenNet, CHGNet, Orb-v3, and MatterSim, against DFT-PBE reference calculations. Test systems include metals (Al, Cu), elemental and compound semiconductors (Si, GaAs, $MoS_2$), binary oxides (MgO, NiO, $TiO_2$), and complex oxides such as $LaCoO_3$ and $La_2NiO_4$. Benchmark results are summarized in **Figure 11**.

***i. Energy Accuracy:***

As shown in **Figure 11a**, all MLPs reproduced DFT energies with chemically reasonable fidelity. Across all systems, most models achieved typical deviations below ~100 meV/atom, consistent with their reported literature performance. Larger deviations were observed for materials lying outside a model's primary training domain (e.g., NiO for CHGNet), emphasizing the importance of selecting an appropriate potential for a given material class.

***ii. Computational Efficiency:***

Computational efficiency results, shown in **Figure 11b**, reveal that MLPs achieve dramatic speed advantages over DFT across system sizes ranging from 8 to 1024 atoms in binary Cu-Mg supercells. All evaluated models exhibit approximately logarithmic scaling with atom count and consistently deliver $10^3$-$10^4\times$ speedups relative to VASP-based DFT calculations, which incur rapidly increasing computational cost with system size and require hours, sometimes even days, for systems exceeding ~128 atoms. Among the models tested, MatterSim provides the fastest inference, followed by CHGNet, Orb-v3, and SevenNet, though all remain orders of magnitude faster than first-principles methods. These findings demonstrate that MLPs offer an extremely efficient alternative for rapid screening, large-supercell exploration (hundreds to thousands of atoms), and MD simulations in scenarios where DFT would be computationally prohibitive.

### 4.3 Case Studies

To further demonstrate Masgent's reliability and practical effectiveness across diverse simulation workflows, we present several representative case studies highlighting convergence analysis, EOS fitting, AIMD simulations, NEB studies, elastic property evaluation, and lightweight ML modeling. Together, these examples highlight how Masgent not only accelerates traditionally labor-intensive tasks, such as structure generation, input preparation, workflow construction, and results analysis, but also ensures that each step adheres to consistent, validated, and reproducible standards. By automatically generating schema-validated inputs, enforcing best-practice defaults, and organizing workflows into robust directory structures, Masgent enables users to perform high-fidelity simulations with confidence and minimal manual intervention.

#### ***i. Case Study 1: Convergence Testing for Aluminum***

As shown in **Figure 12a**, Masgent was used to perform a systematic convergence study for bulk Al by varying the ENCUT values. With a single user request, the AI agent automatically generated all required input files, producing a series of calculations spanning ENCUT values from 300 to 700 eV. Analysis of the resulting energy difference identified convergence within the threshold of 1 meV/atom at ENCUT of approximately 400 eV. This case demonstrates not only Masgent's efficiency, but also its reliability in executing best-practice convergence protocols that yield trustworthy and reproducible results.

***ii. Case Study 2: EOS for Silicon***

**Figure 12b** shows an EOS calculation for bulk Si prepared entirely through Masgent. The AI agent generated a series of volume-scaled structures around the equilibrium configuration and assembled all necessary VASP input files. The resulting energy-volume data were automatically fitted to the Birch-Murnaghan EOS, yielding an equilibrium volume of 20.448 Å$^3$/atom and lattice parameters a = b = c = 5.469 Å, $\alpha = \beta = \gamma = 90°$ at 0 K, in excellent agreement with established experimental and simulation results [54, 55]. This case demonstrates Masgent's ability to produce accurate, publication-quality EOS analyses through a fully automated and reproducible workflow.

***iii. Case Study 3: AIMD for $Li_3InCl_6$***

To demonstrate Masgent's capability in handling finite-temperature dynamical simulations and transport analysis, AIMD simulations were performed for the solid electrolyte $Li_3InCl_6$. Masgent automatically constructed the AIMD workflow, including time-step configuration and temperature scheduling across multiple simulation points. From the resulting trajectories, the AI agent computed mean squared displacements (MSDs) and extracted Li diffusion coefficients (D) at each temperature. As shown in **Figure 12c**, the temperature dependence of the diffusion coefficients follows Arrhenius behavior. A linear fit to $\log_{10}D$ versus 1000/T yields an activation energy of 273.76 meV, in good agreement with reported values (≈ 0.280 eV) for halide-based solid electrolytes [56]. This case highlights Masgent's ability to automate complex AIMD workflows, perform reliable post-processing of dynamical data, and deliver quantitatively meaningful transport properties with minimal user intervention.

***iv. Case Study 4: NEB for $La_2NiO_4$***

Masgent was used to construct and analyze an NEB pathway for oxygen migration in $La_2NiO_4$. The AI agent interpolated an intermediate image between initial and final configurations and produced a complete NEB workflow with ready-to-submit job scripts. After the calculation, Masgent automatically parsed the outputs, reconstructed the reaction pathway using spline smoothing, and exported the reaction coordinate, image energies, and migration barrier. As shown in **Figure 12d**, the resulting energy profile clearly highlights the transition-state region and yields a migration barrier of $\Delta E$ = 1143.37 meV, which falls within the range of previously reported simulation results (0.46–1.34 eV) [57-59]. This case illustrates Masgent's ability to handle complex, multi-image workflows and deliver fully post-processed NEB analyses with minimal user intervention.

### *v. Case Study 5: Elastic Constants for Copper*

Masgent was also employed to compute the elastic constants of FCC Cu by generating a complete set of symmetry-allowed strained configurations. The AI agent prepared input files for each strain state, organized the workflow, and extracted stress tensors from the completed VASP calculations. Linear elasticity analysis yielded $C_{11}$ = 216.62 GPa, $C_{12}$ = 151.67 GPa, and $C_{44}$ = 105.90 GPa at 0 K. These values follow the expected trends and compare reasonably with established experimental results at room temperature ($C_{11}$ = 168.4 GPa, $C_{12}$ = 121.4 GPa, and $C_{44}$ = 75.4 GPa) [60], accounting for thermal softening effects. This example highlights Masgent's reliability in managing multi-step, tensor-level analyses that typically require substantial scripting, data extraction, and manual post- processing.

### *vi. Case Study 6: ML Modeling of Stability for High-Entropy Al-Co-Cr-Fe-Ni Alloys*

Masgent was also applied to a lightweight ML task aimed at predicting the formation enthalpies ($\Delta H_f$) of high-entropy Al-Co-Cr-Fe-Ni alloys, adapted from an established dataset calculated from DFT [53]. The correlation matrix in **Figure 13a** provides initial insight into descriptor relevance, highlighting a positive correlation between Cr content and $\Delta H_f$, as well as a negative correlation with Al content. As shown in **Figure 13b**, Masgent automatically tuned the model hyperparameters using Optuna-based Bayesian optimization, efficiently identifying high-performing configurations while pruning unpromising trials. A neural network regressor was then trained using Masgent's standardized pipeline. The training and validation loss curves in **Figure 13c** exhibit consistent behavior, demonstrating a well-regularized model. The predicted versus true $\Delta H_f$ values in **Figure 13d** fall close to the ideal diagonal, with final metrics of RMSE ≈ 1.48 eV/atom (training) and RMSE ≈ 4.98 eV/atom (validation), along with $R^2$ > 0.89. This case study highlights Masgent's ability to support rapid, data-driven modeling workflows for alloy design, enabling users to explore feature relationships, optimize hyperparameters, and evaluate predictive accuracy, all within a unified platform that complements its physics-based simulation tools.

## 5. Discussion and Future Work

Masgent demonstrates that AI-assisted automation can substantially streamline the setup, execution, and analysis of materials simulations. By unifying DFT automation, MLP simulations, ML utilities, and natural-language interaction within a single platform, Masgent lowers barriers for new practitioners while accelerating the productivity of expert users. At the same time, the current implementation highlights opportunities for further refinement and expansion.

### 5.1 Strengths and Practical Impact

The integration of structured tool-calling with a natural-language interface represents a notable step forward in computational materials science. Compared with traditional scripting-based approaches, Masgent offers several practical advantages:

- **Dramatically reduced setup time:** Complete simulation workflows can be prepared in seconds to minutes, replacing hours or even days of manual setup and scripting.
- **Lowered technical barriers:** The AI agent enables users with minimal programming or HPC experience to construct advanced simulations reliably.
- **Enhanced reproducibility:** Standardized templates and schema-validated input generation minimize human error and ensure consistent simulation protocols across users and projects.
- **Hybrid DFT + MLP capabilities:** Seamless switching between fast MLPs and DFT enables rapid hypothesis testing and efficient pre-screening before expensive calculations.
- **Unified data-centric analysis tools:** Integrated ML utilities support from-scratch model development and pre-trained model applications for materials science.

### 5.2 Limitations and Future Directions

Despite its advantages, Masgent still faces certain limitations that present clear opportunities for future development:

- **Domain awareness and model specialization:** Masgent currently relies on general-purpose LLMs, and ambiguous or underspecified prompts may still require user clarification. Future versions will focus on domain-aware prompting strategies, fine-tuned materials-science LLMs, and task-specific instruction tuning to improve robustness, accuracy, and accessibility.

- **Responsible scientific control:** Masgent is intentionally designed to avoid autonomous decision-making in areas requiring physical judgment, such as the selection of exchange–correlation functionals, Hubbard U parameters, chemical-potential limits, or convergence settings beyond standardized protocols. These choices remain explicitly user-controlled to ensure scientific validity and reproducibility.

- **Lack of direct job execution:** Although Masgent prepares complete workflows, it does not execute VASP jobs or manage HPC queues. Integrating workflow managers such as FireWorks will enable automated job submission, monitoring, dependency handling, and error recovery at HPC scale.

- **Limited workflow coverage:** While convergence tests, EOS fitting, elasticity calculations, AIMD, and NEB workflows are supported, more advanced routines, such as phonon calculations, defect thermodynamics, charge-state analysis, and band-structure workflows, remain under development. Extending coverage to the workflow library is a key priority for future releases.

- **Limited MLP support:** Current potentials are constrained by the scope of their training datasets, reducing reliability for exotic chemistries or extreme conditions. Incorporating additional models (e.g., M3GNet, MACE) and supporting user-supplied MLPs will broaden chemical coverage and improve transferability.

- **Limited ML model support:** Masgent's ML utilities focus on lightweight models for feature analysis, data augmentation, regression, and a small set of embedded pre-trained predictors. Future enhancements will expand the library to include a broader range of pre-trained models covering additional alloy systems and materials properties, as well as more advanced learning frameworks such as graph neural networks, automated feature engineering, uncertainty quantification, and improved model explainability, to support more sophisticated data-driven research.

## 6. Summary

Masgent provides a unified, AI-driven framework that streamlines computational materials research by integrating automated DFT workflow construction, fast MLP simulations, lightweight ML tools, and natural-language interaction. Through schema-validated tool-calling, modular workflow templates, and intelligent parameter handling, Masgent reduces manual effort and user error while enabling rapid, reproducible, and scalable simulation pipelines. Benchmark tests show that Masgent can prepare intricate workflows within seconds to minutes, reproduce established DFT methodologies, and deliver accurate, low-cost approximations via MLP-based calculations.

While certain capabilities, such as phonon workflows and deeper autonomous reasoning, remain under development, the current platform already offers substantial improvements in usability, efficiency, and accessibility. Looking ahead, Masgent provides a robust foundation for incorporating active learning, autonomous optimization loops, expanded workflow coverage, and deeper integration with HPC workflow managers, positioning it as a next-generation computational assistant capable of accelerating both exploratory and production-level materials simulations.

## Competing Interests

The authors declare no competing interests.

## Data Availability

All data generated or analyzed during this study are included in this published article and its associated online examples, available at https://github.com/IMPDGroup/Masgent/tree/main/examples.

## Code Availability

The code for Masgent, along with detailed instructions for usage, is available at https://github.com/IMPDGroup/Masgent.

## Declaration of Generative AI and AI-assisted technologies in the writing process

During the preparation of this work the author(s) used ChatGPT to improve the clarity, grammar, and overall readability of the English writing. After using this tool/service, the author(s) reviewed and edited the content as needed and take(s) full responsibility for the content of the publication.

## Acknowledgments

The authors acknowledge the Advanced Cyberinfrastructure Coordination Ecosystem: Services & Support (ACCESS) program for computational resources provided under Award No. MAT240062. The authors thank ACCESS for their support, resources, and infrastructure that enabled this research. Masgent builds on the open-source materials ecosystem, including ASE, Pymatgen, Icet, and modern machine learning potentials. We thank the developers of these tools for making advanced materials simulation possible.

## Appendix A: Current Functional Overview

- Density Functional Theory (DFT) Simulations
    - Structure Preparation & Manipulation
        - Generate POSCAR from chemical formula
        - Convert POSCAR coordinates (Direct <-> Cartesian)
        - Convert structure file formats (CIF, POSCAR, XYZ)
        - Generate structures with defects (Vacancies, Substitutions, Interstitials)
        - Generate supercells
        - Generate Special Quasirandom Structures (SQS)
        - Generate surface slabs
        - Generate interface structures
        - Visualize structures
    - VASP Input File Preparation
        - Prepare full VASP input files (INCAR, KPOINTS, POTCAR, POSCAR)
        - Generate INCAR templates
            - MPMetalRelaxSet: suggested for metallic structure relaxation
            - MPRelaxSet: suggested for structure relaxation
            - MPStaticSet: suggested for static calculations
            - MPNonSCFBandSet: suggested for non-self-consistent field calculations (Band structure)
            - MPNonSCFDOSSet: suggested for non-self-consistent field calculations (Density of States)
            - MPMDSet: suggested for molecular dynamics simulations
        - Generate KPOINTS with specified accuracy
        - Generate HPC job submission script
    - Standard VASP Workflow Preparation
        - Convergence test (ENCUT, KPOINTS)
        - Equation of State (EOS)
        - Elastic constants calculations
        - Ab-initio Molecular Dynamics (AIMD)

- Nudged Elastic Band (NEB) calculations
    - Standard VASP Workflow Output Analysis
        - Convergence test analysis
        - Equation of State (EOS) analysis
        - Elastic constants analysis
        - Ab-initio Molecular Dynamics (AIMD) analysis
        - Nudged Elastic Band (NEB) analysis
- Fast Simulations Using Machine Learning Potentials (MLPs)
    - Supported MLPs:
        - SevenNet
        - CHGNet
        - Orb-v3
        - MatterSim
    - Implemented Simulations for all MLP:
        - Single Point Energy Calculation
        - Equation of State (EOS) Calculation
        - Elastic Constants Calculation
        - Molecular Dynamics Simulation (NVT)
    - Simple Machine Learning for Materials Science
        - Data Preparation & Feature Analysis
            - Feature analysis and visualization
            - Dimensionality reduction (if too many features)
            - Data augmentation (if limited data)
        - Model Design & Hyperparameter Tuning
        - Model Training & Evaluation
        - Model Retraining with New Data
        - Pre-trained Models Prediction
            - Mechanical Properties Prediction in Sc-modified Al-Mg-Si Alloys
            - Phase Stability & Elastic Properties Prediction in Al-Co-Cr-Fe-Ni High-Entropy Alloys

# References

[1] K. Gubaev, E.V. Podryabinkin, G.L. Hart, A.V. Shapeev, Accelerating high-throughput searches for new alloys with active learning of interatomic potentials, Computational Materials Science 156 (2019) 148-156.

[2] P. Korotaev, I. Novoselov, A. Yanilkin, A. Shapeev, Accessing thermal conductivity of complex compounds by machine learning interatomic potentials, Physical Review B 100(14) (2019) 144308.

[3] B. Mortazavi, E.V. Podryabinkin, I.S. Novikov, S. Roche, T. Rabczuk, X. Zhuang, A.V. Shapeev, Efficient machine-learning based interatomic potentialsfor exploring thermal conductivity in two-dimensional materials, Journal of Physics: Materials 3(2) (2020) 02LT02.

[4] V. Ladygin, P.Y. Korotaev, A. Yanilkin, A. Shapeev, Lattice dynamics simulation using machine learning interatomic potentials, Computational Materials Science 172 (2020) 109333.

[5] B. Mortazavi, E.V. Podryabinkin, S. Roche, T. Rabczuk, X. Zhuang, A.V. Shapeev, Machine-learning interatomic potentials enable first-principles multiscale modeling of lattice thermal conductivity in graphene/borophene heterostructures, Materials Horizons 7(9) (2020) 2359-2367.

[6] E.V. Podryabinkin, E.V. Tikhonov, A.V. Shapeev, A.R. Oganov, Accelerating crystal structure prediction by machine-learning interatomic potentials with active learning, Physical Review B 99(6) (2019) 064114.

[7] S. Yang, G. Liu, Y. Zhong, Revisit the VEC criterion in high entropy alloys (HEAs) with high-throughput ab initio calculations: A case study with Al-Co-Cr-Fe-Ni system, Journal of Alloys and Compounds (2022) 165477.

[8] S. Yang, G. Liu, Y. Zhong, Ab initio investigations on the electronic properties and stability of Cu-substituted lead apatite (LK-99) family with different doping concentrations (x= 0, 1, 2), Materials Today Communications 37 (2023) 107379.

[9] G. Liu, S. Yang, Y. Zhong, A Computational Framework for Interface Design Using Lattice Matching, Machine Learning Potentials, and Active Learning: A Case Study on $LaCoO_3$/$La_2$NiO4, Materials Today Physics (2025) 101940.

[10] G. Kresse, J. Furthmüller, Efficient iterative schemes for ab initio total-energy calculations using a plane-wave basis set, Physical Review B 54(16) (1996) 11169.

[11] G. Kresse, J. Furthmüller, Efficiency of ab-initio total energy calculations for metals and semiconductors using a plane-wave basis set, Computational Materials Science 6(1) (1996) 15-50.

[12] S.P. Ong, W.D. Richards, A. Jain, G. Hautier, M. Kocher, S. Cholia, D. Gunter, V.L. Chevrier, K.A. Persson, G. Ceder, Python Materials Genomics (pymatgen): A robust, open-source python library for materials analysis, Computational Materials Science 68 (2013) 314-319.

[13] A.H. Larsen, J.J. Mortensen, J. Blomqvist, I.E. Castelli, R. Christensen, M. Dułak, J. Friis, M.N. Groves, B. Hammer, C. Hargus, The atomic simulation environment—a Python library for working with atoms, Journal of Physics: Condensed Matter 29(27) (2017) 273002.

[14] V. Wang, N. Xu, J.-C. Liu, G. Tang, W.-T. Geng, VASPKIT: A user-friendly interface facilitating high-throughput computing and analysis using VASP code, Computer Physics Communications 267 (2021) 108033.

[15] W. Yi, G. Tang, X. Chen, B. Yang, X. Liu, qvasp: A flexible toolkit for VASP users in materials simulations, Computer Physics Communications 257 (2020) 107535.

[16] K. Mathew, J.H. Montoya, A. Faghaninia, S. Dwarakanath, M. Aykol, H. Tang, I.-h. Chu, T. Smidt, B. Bocklund, M. Horton, Atomate: A high-level interface to generate, execute, and analyze computational materials science workflows, Computational Materials Science 139 (2017) 140-152.

[17] A.M. Ganose, H. Sahasrabuddhe, M. Asta, K. Beck, T. Biswas, A. Bonkowski, J. Bustamante, X. Chen, Y. Chiang, D.C. Chrzan, Atomate2: Modular workflows for materials science, Digital Discovery (2025).
[18] F. Pedregosa, G. Varoquaux, A. Gramfort, V. Michel, B. Thirion, O. Grisel, M. Blondel, P. Prettenhofer, R. Weiss, V. Dubourg, Scikit-learn: Machine learning in Python, the Journal of machine Learning research 12 (2011) 2825-2830.
[19] M. Abadi, P. Barham, J. Chen, Z. Chen, A. Davis, J. Dean, M. Devin, S. Ghemawat, G. Irving, M. Isard, TensorFlow: a system for large-scale machine learning, 12th USENIX symposium on operating systems design and implementation (OSDI 16), 2016, pp. 265-283.
[20] A. Paszke, S. Gross, F. Massa, A. Lerer, J. Bradbury, G. Chanan, T. Killeen, Z. Lin, N. Gimelshein, L. Antiga, PyTorch: An imperative style, high-performance deep learning library, Advances in neural information processing systems 32 (2019).
[21] A. Jain, S.P. Ong, G. Hautier, W. Chen, W.D. Richards, S. Dacek, S. Cholia, D. Gunter, D. Skinner, G. Ceder, Commentary: The Materials Project: A materials genome approach to accelerating materials innovation, APL materials 1(1) (2013).
[22] S. Kirklin, J.E. Saal, B. Meredig, A. Thompson, J.W. Doak, M. Aykol, S. Rühl, C. Wolverton, The Open Quantum Materials Database (OQMD): assessing the accuracy of DFT formation energies, npj Computational Materials 1(1) (2015) 1-15.
[23] J.E. Saal, S. Kirklin, M. Aykol, B. Meredig, C. Wolverton, Materials design and discovery with high-throughput density functional theory: the open quantum materials database (OQMD), Jom 65 (2013) 1501-1509.
[24] Y. Park, J. Kim, S. Hwang, S. Han, Scalable parallel algorithm for graph neural network interatomic potentials in molecular dynamics simulations, Journal of chemical theory and computation 20(11) (2024) 4857-4868.
[25] B. Deng, P. Zhong, K. Jun, J. Riebesell, K. Han, C.J. Bartel, G. Ceder, CHGNet as a pretrained universal neural network potential for charge-informed atomistic modelling, Nature Machine Intelligence 5(9) (2023) 1031-1041.
[26] M. Neumann, J. Gin, B. Rhodes, S. Bennett, Z. Li, H. Choubisa, A. Hussey, J. Godwin, Orb: A fast, scalable neural network potential, arXiv preprint arXiv:2410.22570 (2024).
[27] B. Rhodes, S. Vandenhaute, V. Šimkus, J. Gin, J. Godwin, T. Duignan, M. Neumann, Orb-v3: atomistic simulation at scale, arXiv preprint arXiv:2504.06231 (2025).
[28] H. Yang, C. Hu, Y. Zhou, X. Liu, Y. Shi, J. Li, G. Li, Z. Chen, S. Chen, C. Zeni, MatterSim: A deep learning atomistic model across elements, temperatures and pressures, arXiv preprint arXiv:2405.04967 (2024).
[29] C. Chen, S.P. Ong, A universal graph deep learning interatomic potential for the periodic table, Nature Computational Science 2(11) (2022) 718-728.
[30] I. Batatia, D.P. Kovacs, G. Simm, C. Ortner, G. Csányi, MACE: Higher order equivariant message passing neural networks for fast and accurate force fields, Advances in neural information processing systems 35 (2022) 11423-11436.
[31] S. Takamoto, C. Shinagawa, D. Motoki, K. Nakago, W. Li, I. Kurata, T. Watanabe, Y. Yayama, H. Iriguchi, Y. Asano, Towards universal neural network potential for material discovery applicable to arbitrary combination of 45 elements, Nature Communications 13(1) (2022) 2991.
[32] A. Vaswani, N. Shazeer, N. Parmar, J. Uszkoreit, L. Jones, A.N. Gomez, Ł. Kaiser, I. Polosukhin, Attention is all you need, Advances in neural information processing systems 30 (2017).

[33] T. Brown, B. Mann, N. Ryder, M. Subbiah, J.D. Kaplan, P. Dhariwal, A. Neelakantan, P. Shyam, G. Sastry, A. Askell, Language models are few-shot learners, Advances in neural information processing systems 33 (2020) 1877-1901.
[34] J. Kaplan, S. McCandlish, T. Henighan, T.B. Brown, B. Chess, R. Child, S. Gray, A. Radford, J. Wu, D. Amodei, Scaling laws for neural language models, arXiv preprint arXiv:2001.08361 (2020).
[35] T. Schick, J. Dwivedi-Yu, R. Dessì, R. Raileanu, M. Lomeli, E. Hambro, L. Zettlemoyer, N. Cancedda, T. Scialom, Toolformer: Language models can teach themselves to use tools, Advances in Neural Information Processing Systems 36 (2023) 68539-68551.
[36] H. Yang, S. Yue, Y. He, Auto-gpt for online decision making: Benchmarks and additional opinions, arXiv preprint arXiv:2306.02224 (2023).
[37] S. Yao, J. Zhao, D. Yu, N. Du, I. Shafran, K.R. Narasimhan, Y. Cao, React: Synergizing reasoning and acting in language models, The eleventh international conference on learning representations, 2022.
[38] G. Wang, Y. Xie, Y. Jiang, A. Mandlekar, C. Xiao, Y. Zhu, L. Fan, A. Anandkumar, Voyager: An open-ended embodied agent with large language models, arXiv preprint arXiv:2305.16291 (2023).
[39] A. Maćkiewicz, W. Ratajczak, Principal components analysis (PCA), Computers & Geosciences 19(3) (1993) 303-342.
[40] C. Doersch, Tutorial on variational autoencoders, arXiv preprint arXiv:1606.05908 (2016).
[41] D.P. Kingma, M. Welling, An introduction to variational autoencoders, Foundations and Trends® in Machine Learning 12(4) (2019) 307-392.
[42] X. Gong, B. Tang, R. Zhu, W. Liao, L. Song, Data augmentation for electricity theft detection using conditional variational auto-encoder, Energies 13(17) (2020) 4291.
[43] T. Akiba, S. Sano, T. Yanase, T. Ohta, M. Koyama, Optuna: A next-generation hyperparameter optimization framework, Proceedings of the 25th ACM SIGKDD International Conference on Knowledge Discovery & Data Mining, 2019, pp. 2623-2631.
[44] N. Rego, D. Koes, 3Dmol.js: molecular visualization with WebGL, Bioinformatics 31(8) (2015) 1322-1324.
[45] K. Momma, F. Izumi, VESTA 3 for three-dimensional visualization of crystal, volumetric and morphology data, Applied Crystallography 44(6) (2011) 1272-1276.
[46] K. Momma, F. Izumi, VESTA: a three-dimensional visualization system for electronic and structural analysis, Applied Crystallography 41(3) (2008) 653-658.
[47] J.-X. Shen, J. Varley, pymatgen-analysis-defects: A Python package for analyzing point defects in crystalline materials, Journal of Open Source Software 9(93) (2024) 5941.
[48] M. Ångqvist, W.A. Muñoz, J.M. Rahm, E. Fransson, C. Durniak, P. Rozyczko, T.H. Rod, P. Erhart, ICET–a Python library for constructing and sampling alloy cluster expansions, Advanced Theory and Simulations 2(7) (2019) 1900015.
[49] A. Zur, T. McGill, Lattice match: An application to heteroepitaxy, Journal of applied physics 55(2) (1984) 378-386.
[50] S. Watanabe, Tree-structured parzen estimator: Understanding its algorithm components and their roles for better empirical performance, arXiv preprint arXiv:2304.11127 (2023).
[51] J. Gao, J. Zhong, G. Liu, S. Zhang, J. Zhang, Z. Liu, B. Song, L. Zhang, Accelerated discovery of high-performance Al-Si-Mg-Sc casting alloys by integrating active learning with high-throughput CALPHAD calculations, Science and Technology of Advanced Materials 24(1) (2023) 2196242.

[52] G. Liu, J. Gao, C. Che, Z. Lu, W. Yi, L. Zhang, Optimization of casting means and heat treatment routines for improving mechanical and corrosion resistance properties of A356-0.54 Sc casting alloy, Materials Today Communications 24 (2020) 101227.
[53] G. Liu, S. Yang, Y. Zhong, High-Entropy Materials Design by Integrating the First-Principles Calculations and Machine Learning: A Case Study in the Al-Co-Cr-Fe-Ni System, High Entropy Alloys & Materials (2024) 1-14.
[54] Materials Data on Si by Materials Project, United States, 2020.
[55] W. Bond, W. Kaiser, Interstitial versus substitutional oxygen in silicon, Journal of Physics and Chemistry of Solids 16(1-2) (1960) 44-45.
[56] C. Zhang, H. Zhuang, Z. Qi, X. Liu, Y. Ren, The effect of defects for the ion transport of $Li_3ScCl_6$ and $Li_3InCl_6$ with the interface of lithium metal anode: A first-principles study, Materials Today Communications 35 (2023) 105764.
[57] Z. Du, Z. Zhang, A. Niemczyk, A. Olszewska, N. Chen, K. Świerczek, H. Zhao, Unveiling the effects of A-site substitutions on the oxygen ion migration in $A_{2-x}A'_xNiO_{4+\delta}$ by first principles calculations, Physical Chemistry Chemical Physics 20(33) (2018) 21685-21692.
[58] S. Yang, G. Liu, Y.-L. Lee, J.-M. Bassat, J. Gamon, A. Villesuzanne, J. Pietras, X.-D. Zhou, Y. Zhong, A systematic ab initio study of vacancy formation energy, diffusivity, and ionic conductivity of $Ln_2NiO_{4+\delta}$ (Ln= La, Nd, Pr), Journal of Power Sources 576 (2023) 233200.
[59] S. Yang, G. Liu, W. Li, E.M. Sabolsky, X. Liu, Y. Zhong, Ab initio study on the effect of A-site doping on the stability, equilibrium volume, activation energy barrier, and oxygen diffusivity in La2-xAxNiO4+δ, International Journal of Hydrogen Energy 119 (2025) 239-251.
[60] H. Ledbetter, E. Naimon, Elastic properties of metals and alloys. II. Copper, Journal of physical and chemical reference data 3(4) (1974) 897-935.

## Tables

*Table 1 Comparison of Masgent with existing materials simulation tools.*

| Capability | ASE | Pymatgen | VASPkit, Atomate, qvasp, etc. | MLPs (SevenNet, CHGNet, Orb-v3, MatterSim, etc.) | Masgent |
|---|---|---|---|---|---|
| Structure manipulation | Yes | Yes | Yes | No | Yes |
| VASP inputs | No | Limited | Yes | No | Yes |
| Workflow templates | No | No | Yes | No | Yes |
| MLP simulations | Limited | No | No | Yes | Yes (Unified API) |
| AI agent | No | No | No | No | Yes |
| ML utilities | No | No | No | No | Yes |

*Note: Masgent's distinction lies not in individual capabilities, but in their unified, AI-assisted orchestration and schema-validated execution.*

*Table 2 Overview of large language models (LLMs) supported by Masgent.*

| Provider | Model | API Key Required |
| --- | --- | --- |
| Masgent | Masgent AI | No |
| OpenAI | GPT-5 Nano | Yes |
| Anthropic | Claude Sonnet 4.5 | Yes |
| Google | Gemini 2.5 Flash | Yes |
| xAI | Grok 4.1 Fast | Yes |
| DeepSeek | DeepSeek Chat | Yes |
| Alibaba | Qwen Flash | Yes |

*Table 3 Default atomic radii used in Masgent for structure visualization.*

| Element | Radius | Element | Radius | Element | Radius | Element | Radius | Element | Radius |
|---|---|---|---|---|---|---|---|---|---|
| H | 0.46 | Sc | 1.64 | Nb | 1.47 | Pm | 1.81 | Tl | 1.71 |
| He | 1.22 | Ti | 1.47 | Mo | 1.40 | Sm | 1.81 | Pb | 1.75 |
| Li | 1.57 | V | 1.35 | Tc | 1.35 | Eu | 2.06 | Bi | 1.82 |
| Be | 1.12 | Cr | 1.29 | Ru | 1.34 | Gd | 1.79 | Po | 1.77 |
| B | 0.81 | Mn | 1.37 | Rh | 1.34 | Tb | 1.77 | At | 0.62 |
| C | 0.77 | Fe | 1.26 | Pd | 1.37 | Dy | 1.77 | Fr | 1.00 |
| N | 0.74 | Co | 1.25 | Ag | 1.44 | Ho | 1.76 | Ra | 2.35 |
| O | 0.74 | Ni | 1.25 | Cd | 1.52 | Er | 1.75 | Ac | 2.03 |
| F | 0.72 | Cu | 1.28 | In | 1.67 | Tm | 1.00 | Th | 1.80 |
| Ne | 1.60 | Zn | 1.37 | Sn | 1.58 | Yb | 1.94 | Pa | 1.63 |
| Na | 1.91 | Ga | 1.53 | Sb | 1.41 | Lu | 1.72 | U | 1.56 |
| Mg | 1.60 | Ge | 1.22 | Te | 1.37 | Hf | 1.59 | Np | 1.56 |
| Al | 1.43 | As | 1.21 | I | 1.33 | Ta | 1.47 | Pu | 1.64 |
| Si | 1.18 | Se | 1.04 | Xe | 2.18 | W | 1.41 | Am | 1.73 |
| P | 1.10 | Br | 1.14 | Cs | 2.71 | Re | 1.37 | Others | 0.80 |
| S | 1.04 | Kr | 1.98 | Ba | 2.24 | Os | 1.35 | | |
| Cl | 0.99 | Rb | 2.50 | La | 1.88 | Ir | 1.36 | | |
| Ar | 1.92 | Sr | 2.15 | Ce | 1.82 | Pt | 1.39 | | |
| K | 2.35 | Y | 1.82 | Pr | 1.82 | Au | 1.44 | | |
| Ca | 1.97 | Zr | 1.60 | Nd | 1.82 | Hg | 1.55 | | |

## Figures

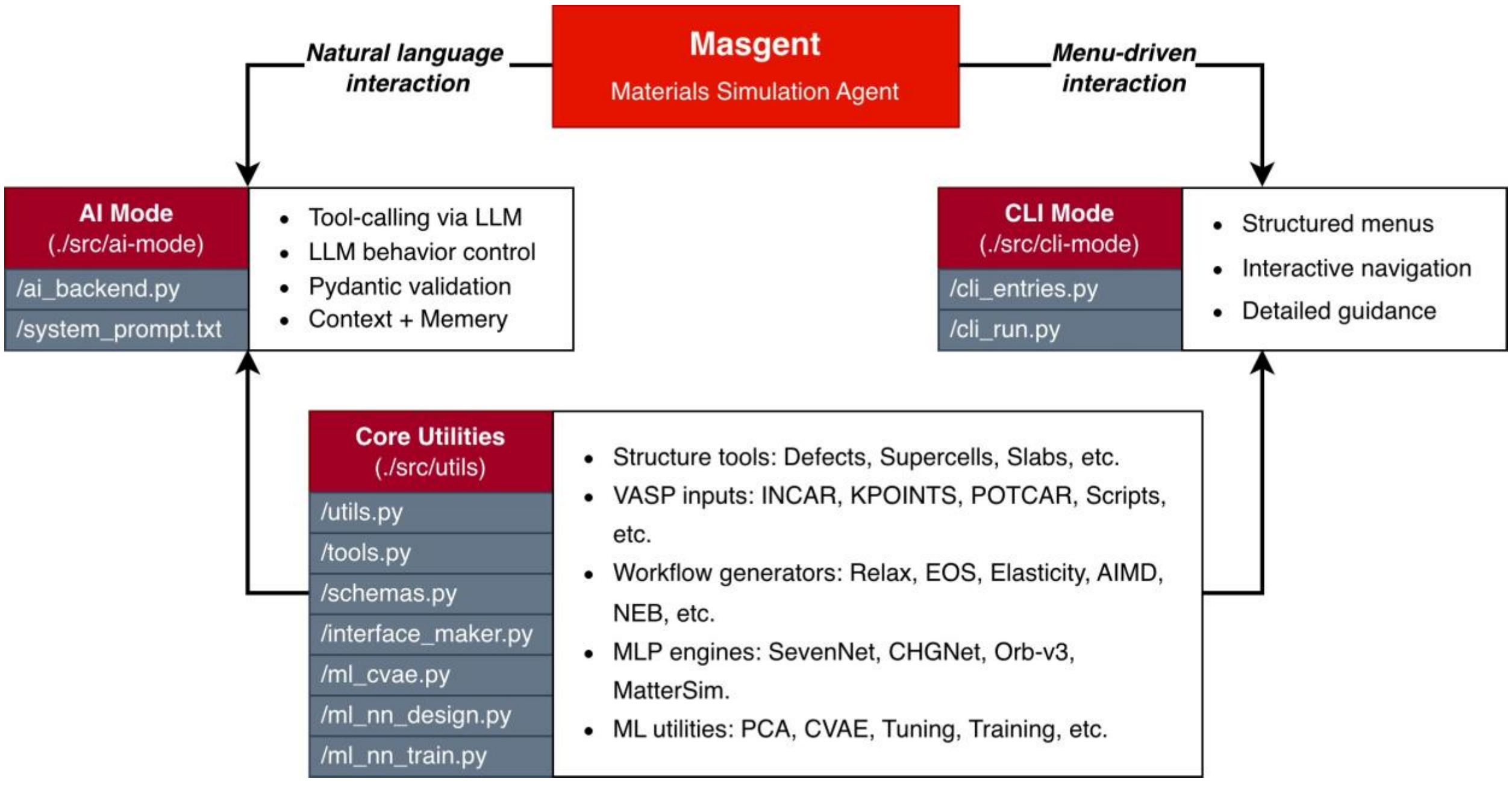

*Figure 1 Schematic overview of Masgent's modular architecture. The AI Mode and CLI Mode provide two user interfaces that both access the Core Utilities layer, which contains all computational tools for structure operations, VASP input generation, workflow automation, MLP simulations, and ML utilities.*

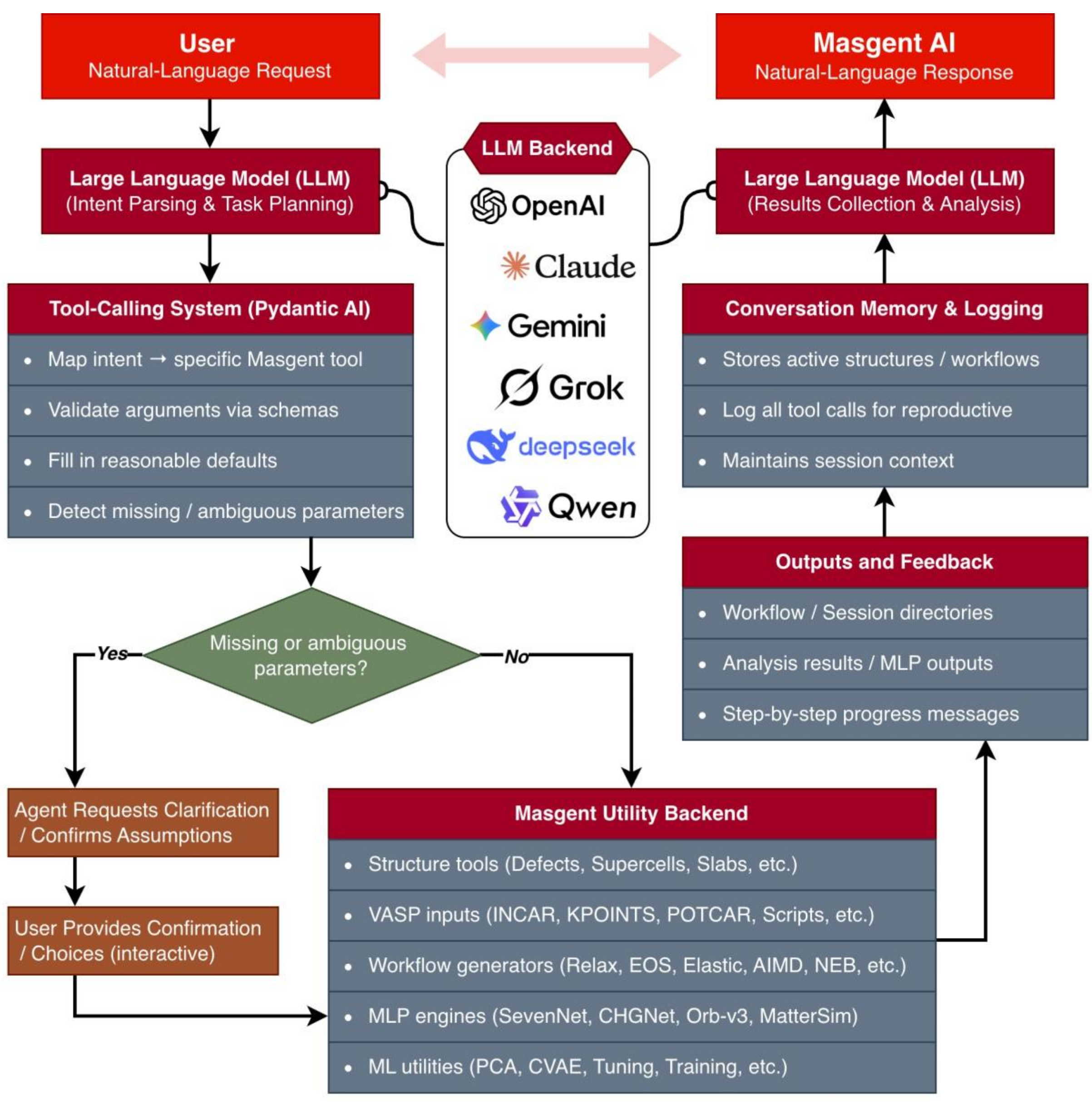

*Figure 2 AI agent workflow of Masgent. User requests are parsed by an LLM and routed through a schema-validated tool-calling system. The agent confirms assumptions, requests clarification when needed, and generates step-by-step feedback during execution. Validated operations are executed by Masgent's computational backend, and all actions are stored in conversation memory for context-aware, multi-turn interactions.*

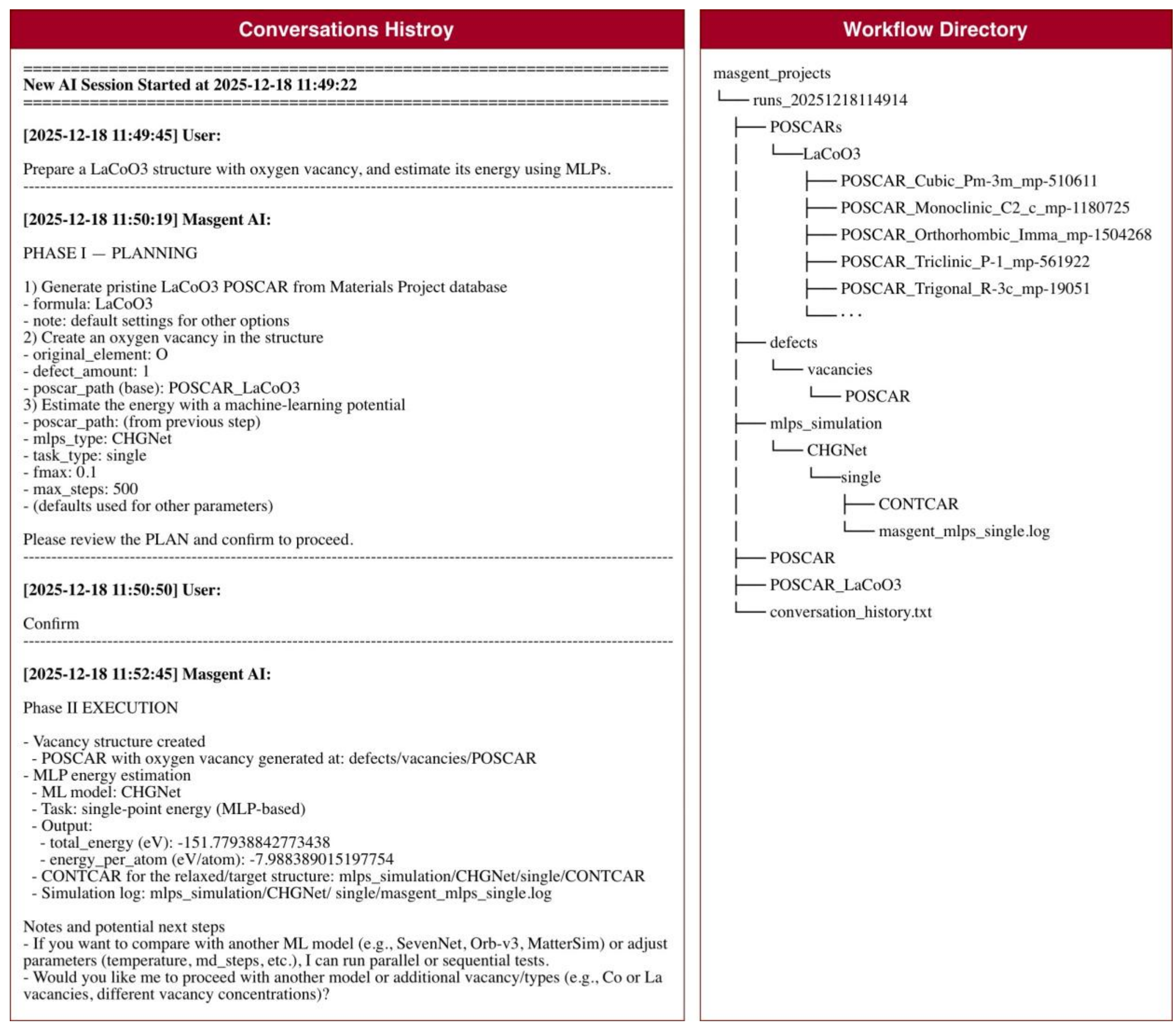

*Figure 3 End-to-end example of Masgent's AI agent generating a complete workflow from a single natural-language request. (Left) The agent interacts with the user via clarifying questions, confirmation steps, and progress updates to build a valid simulation workflow. (Right) The resulting directory structure includes the fetched crystal structure, vacancy defect, and MLP energy evaluation.*

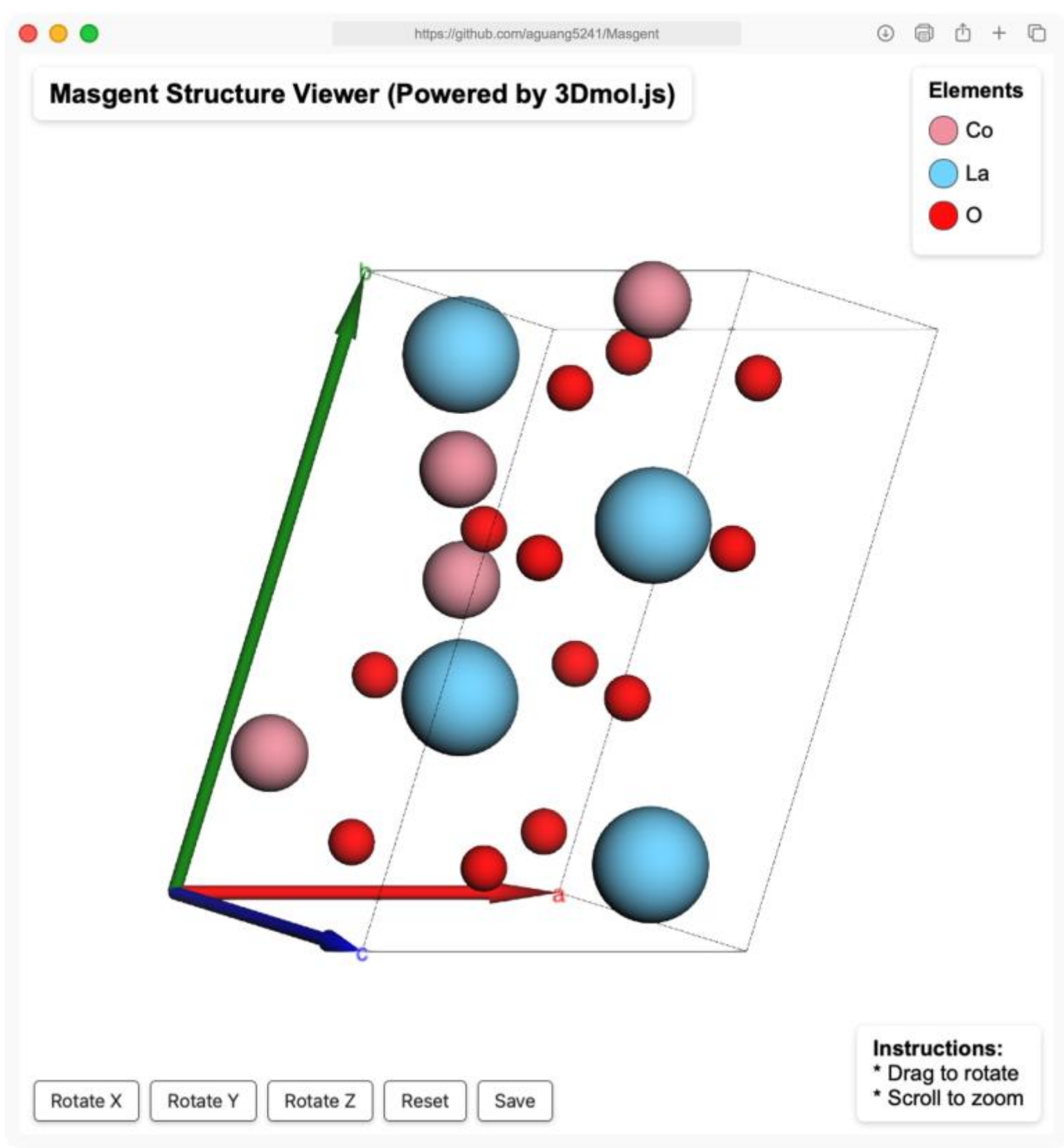

*Figure 4 Interactive crystal structure visualization in Masgent. The viewer supports rotation and zooming, displays the unit cell and lattice vectors, and uses VESTA-consistent atomic radii for standardized visualization.*

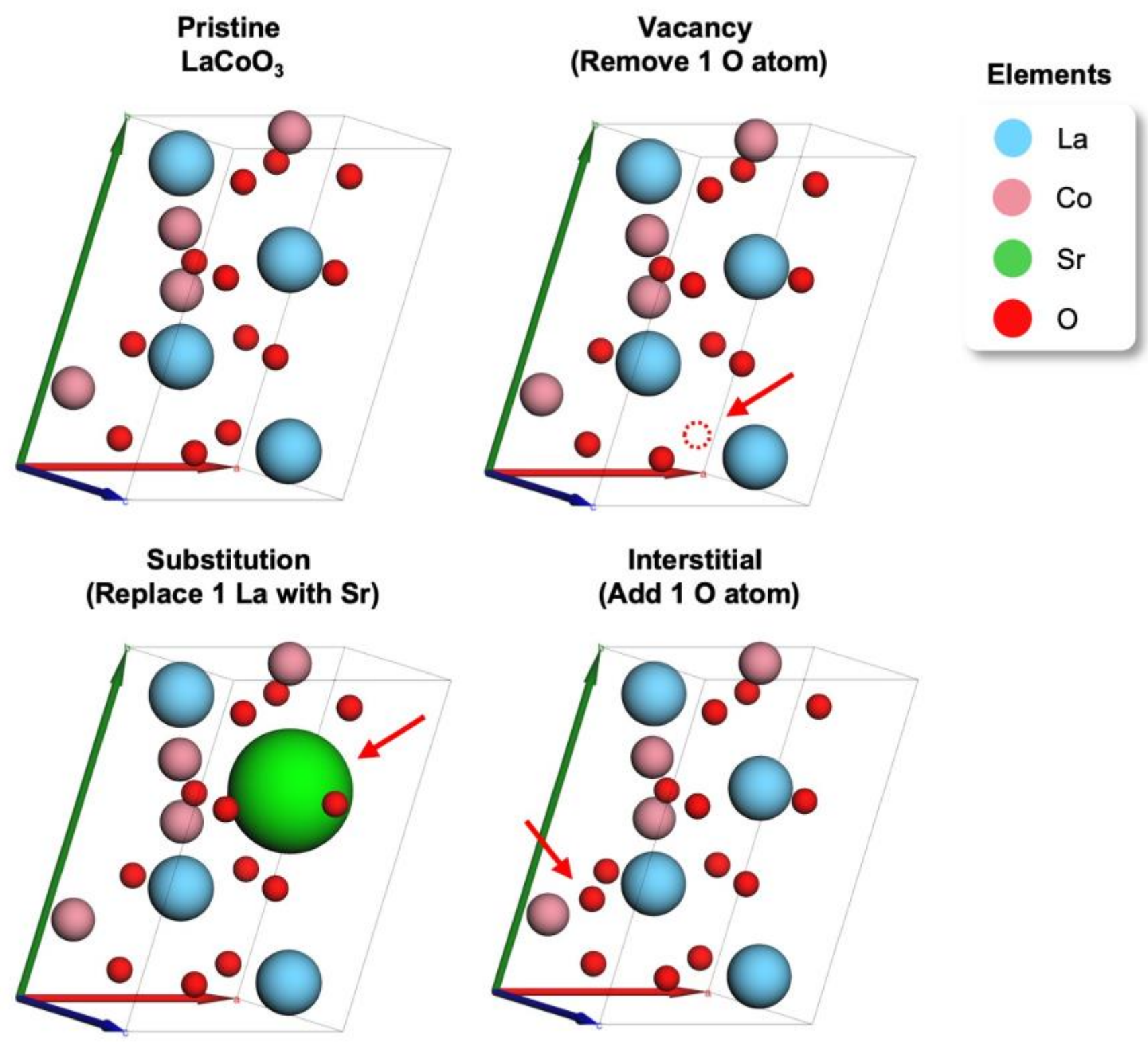

*Figure 5 Defect generation in Masgent, illustrating pristine $LaCoO_3$ and representative vacancy, substitution, and interstitial defect configurations generated from the parent structure.*

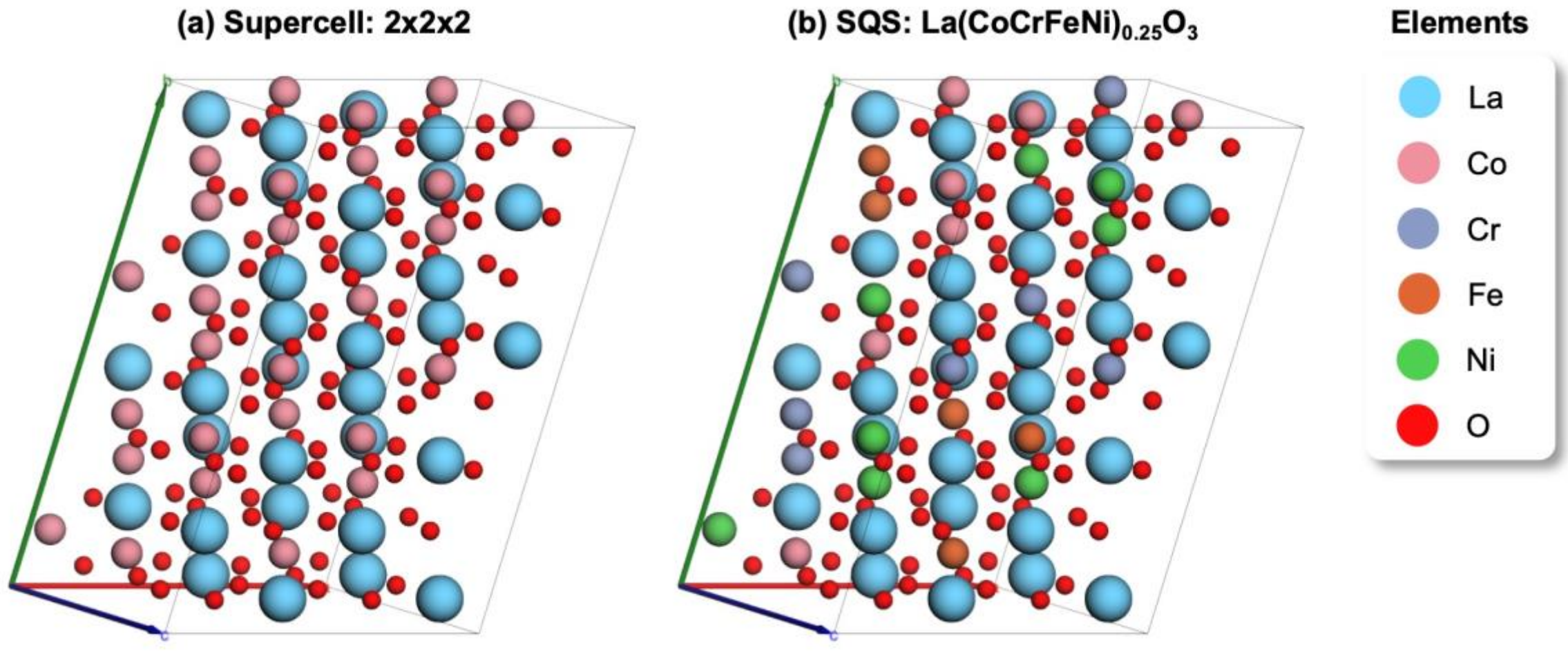

*Figure 6 Supercell and SQS generation in Masgent, showing (a) a 2×2×2 supercell constructed from the primitive $LaCoO_3$ cell and (b) an SQS generated using the Icet package to model compositional disorder.*

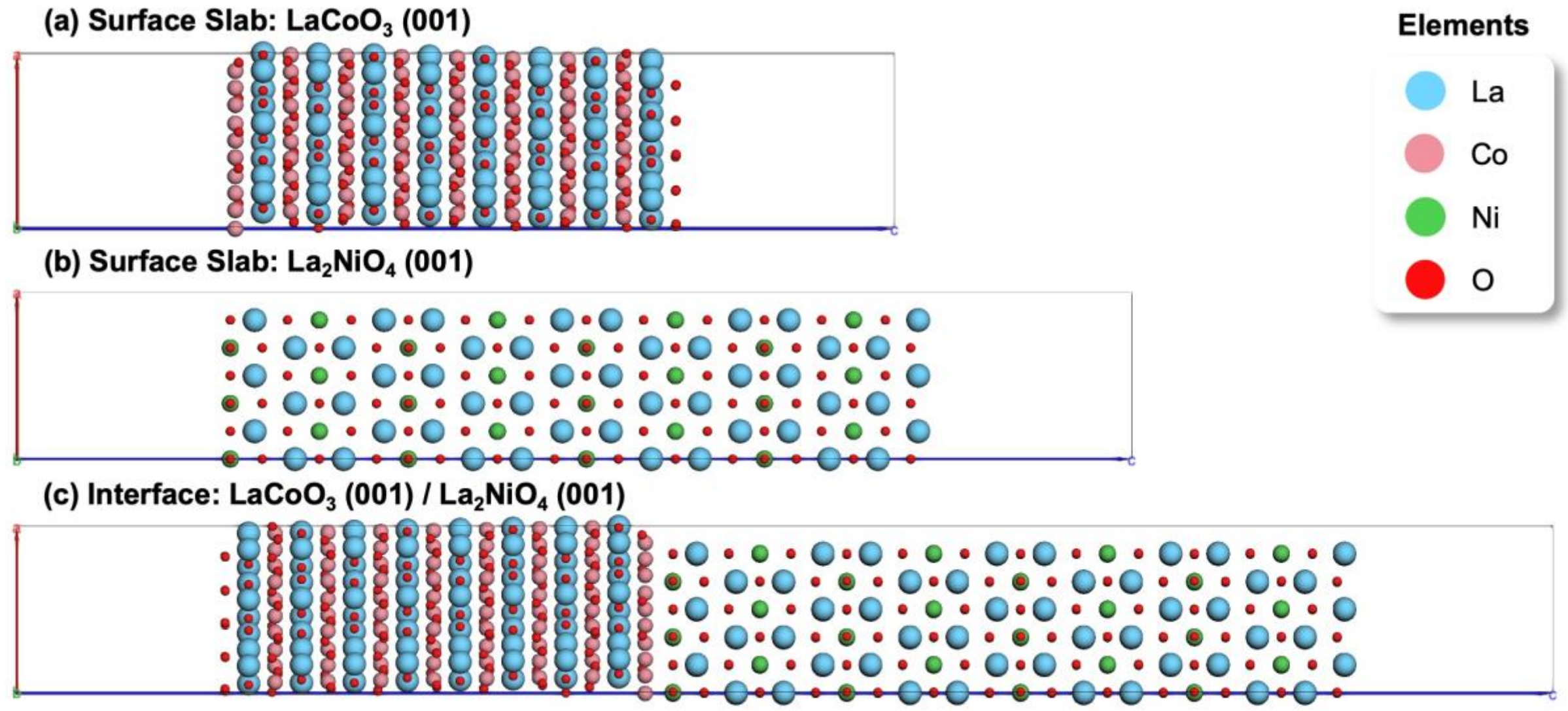

*Figure 7 Surface slab and interface construction in Masgent, showing (a) $LaCoO_3$ (001) and (b) $La_2NiO_4$ (001) surface slabs generated with specified slab thickness and vacuum spacing, and (c) a $LaCoO_3$ (001)/$La_2NiO_4$ (001) interface constructed using lattice-matching algorithms.*

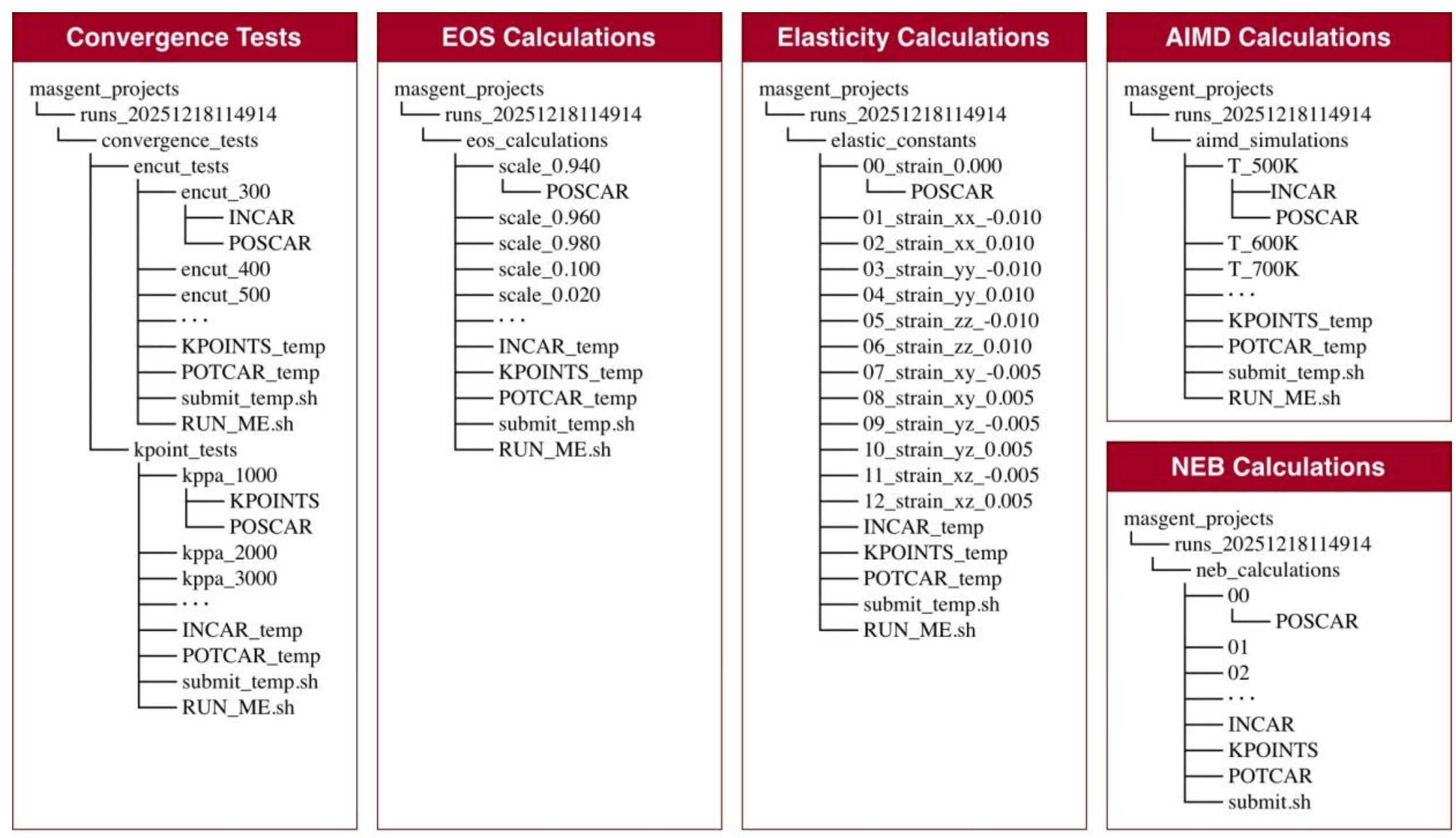

*Figure 8 Example VASP workflow structure generated by Masgent, illustrating automatically organized directory trees for convergence tests, EOS, elasticity, AIMD, and NEB calculations within a unified and reproducible DFT simulation pipeline.*

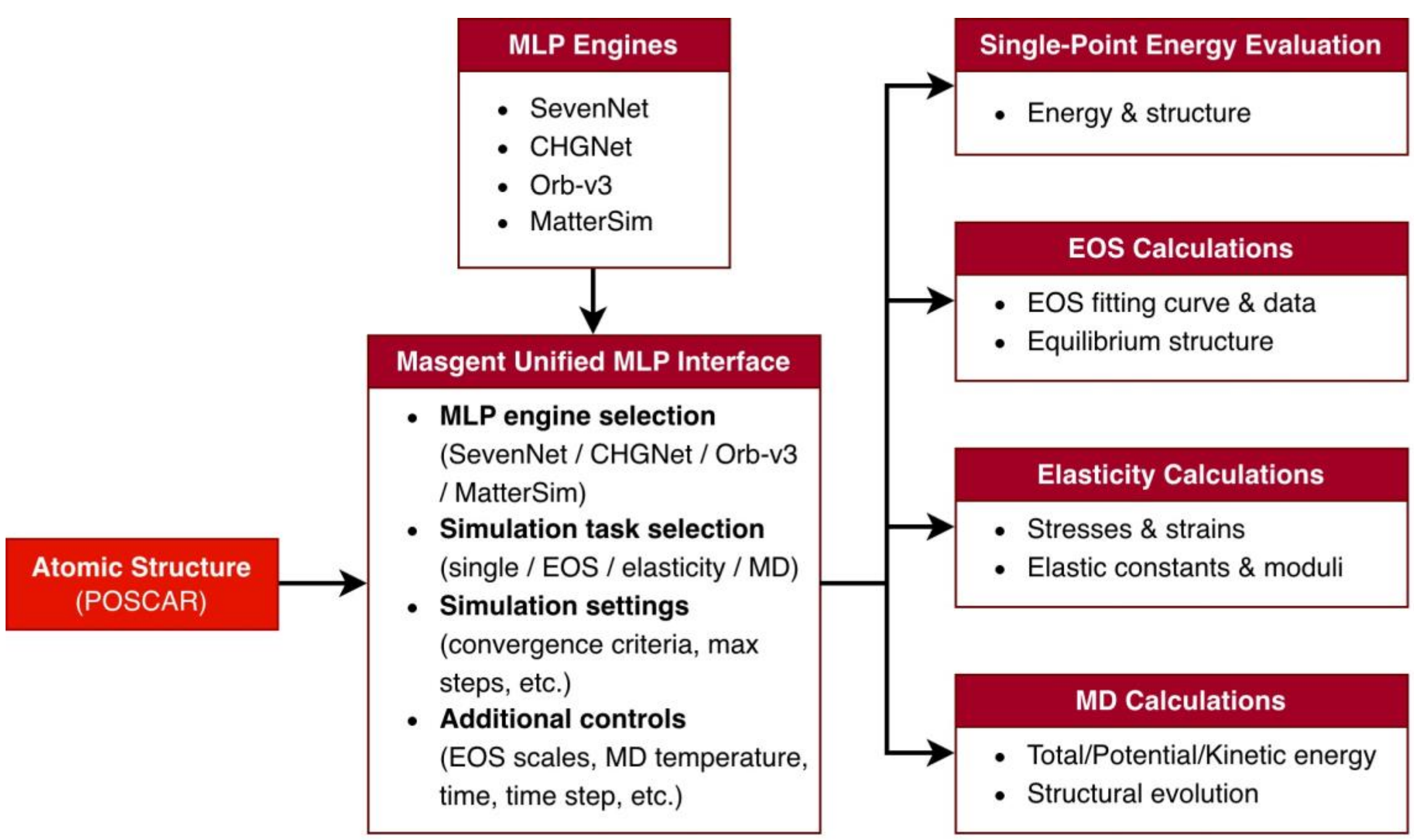

*Figure 9 Unified MLP framework in Masgent. Atomic structures are processed through a unified MLP interface that provides consistent access to multiple engines (SevenNet, CHGNet, Orb-v3, and MatterSim), enabling standardized single-point energy, EOS, elastic constant, and ab-initio molecular dynamics calculations.*

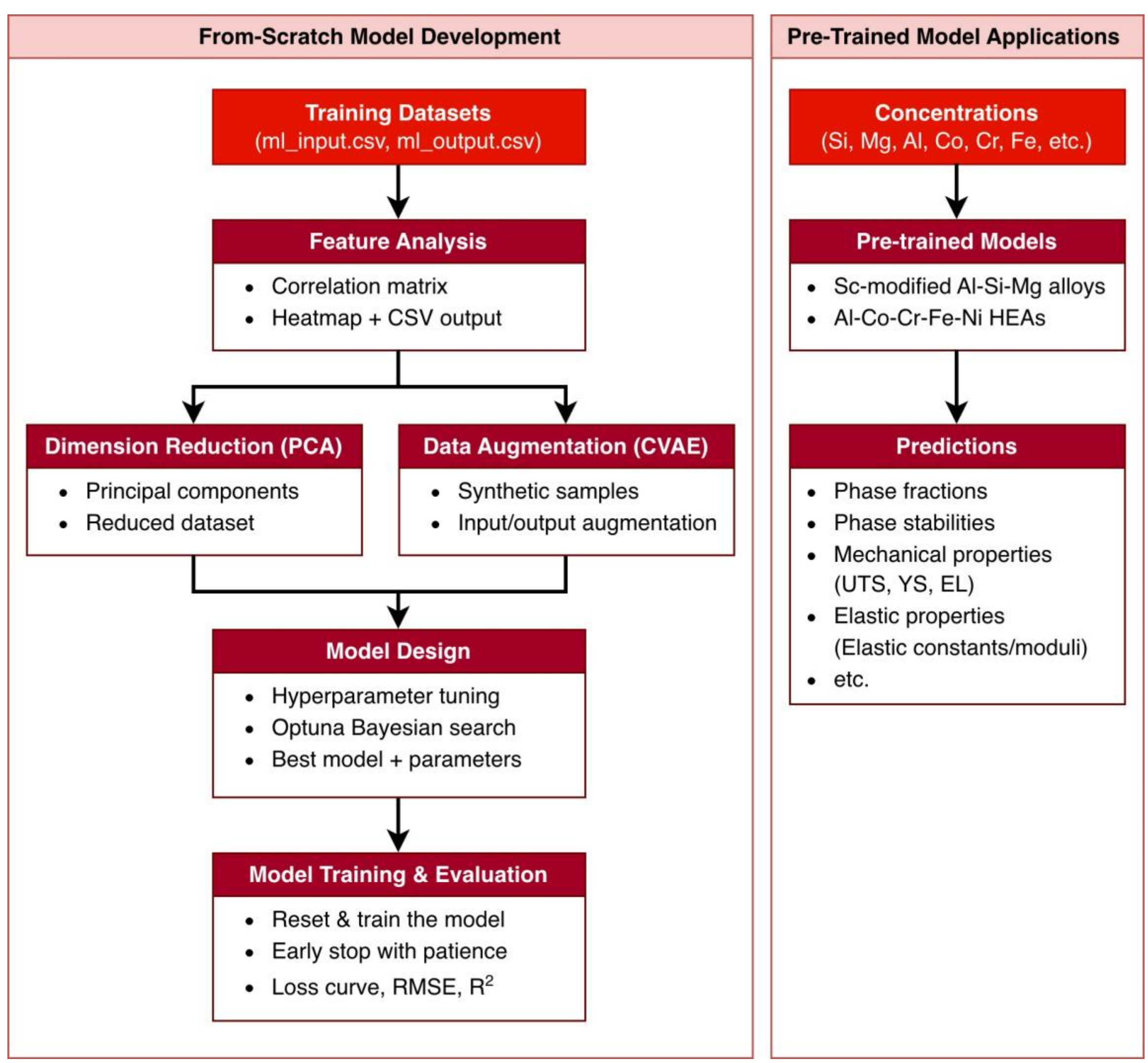

*Figure 10 Overview of lightweight ML utilities in Masgent, illustrating two workflows: from-scratch model development (left), including feature analysis, PCA-based dimensionality reduction, CVAE-based data augmentation, hyperparameter optimization, and model training/evaluation; and pre-trained model applications (right), enabling property prediction from alloy compositions using existing models.*

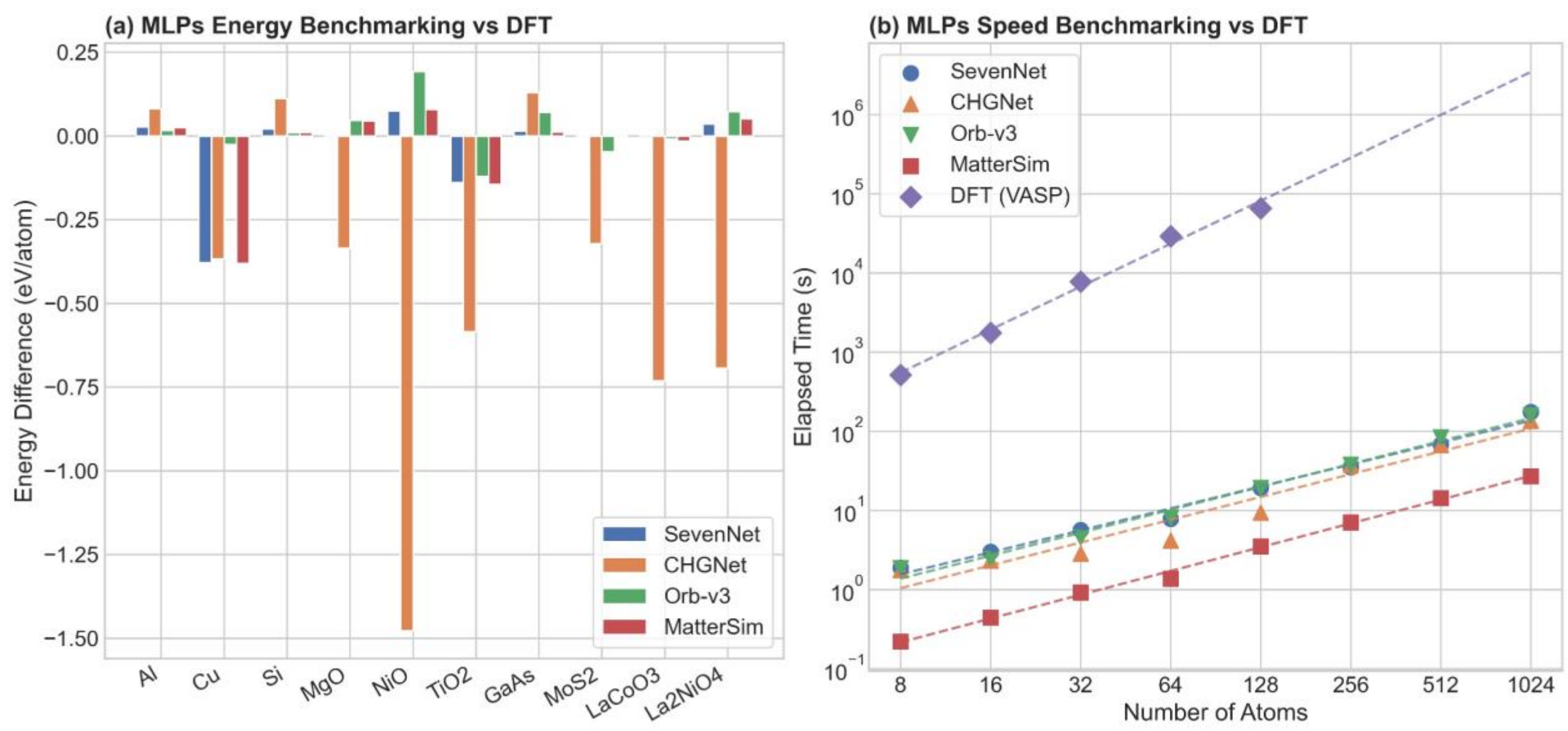

*Figure 11 Performance benchmarking of Masgent's MLPs. (a) Single-point energy errors of SevenNet, CHGNet, Orb-v3, and MatterSim compared to DFT across diverse materials. (b) Computational scaling showing $10^3$-$10^4$ times speedups over DFT for increasing system sizes.*

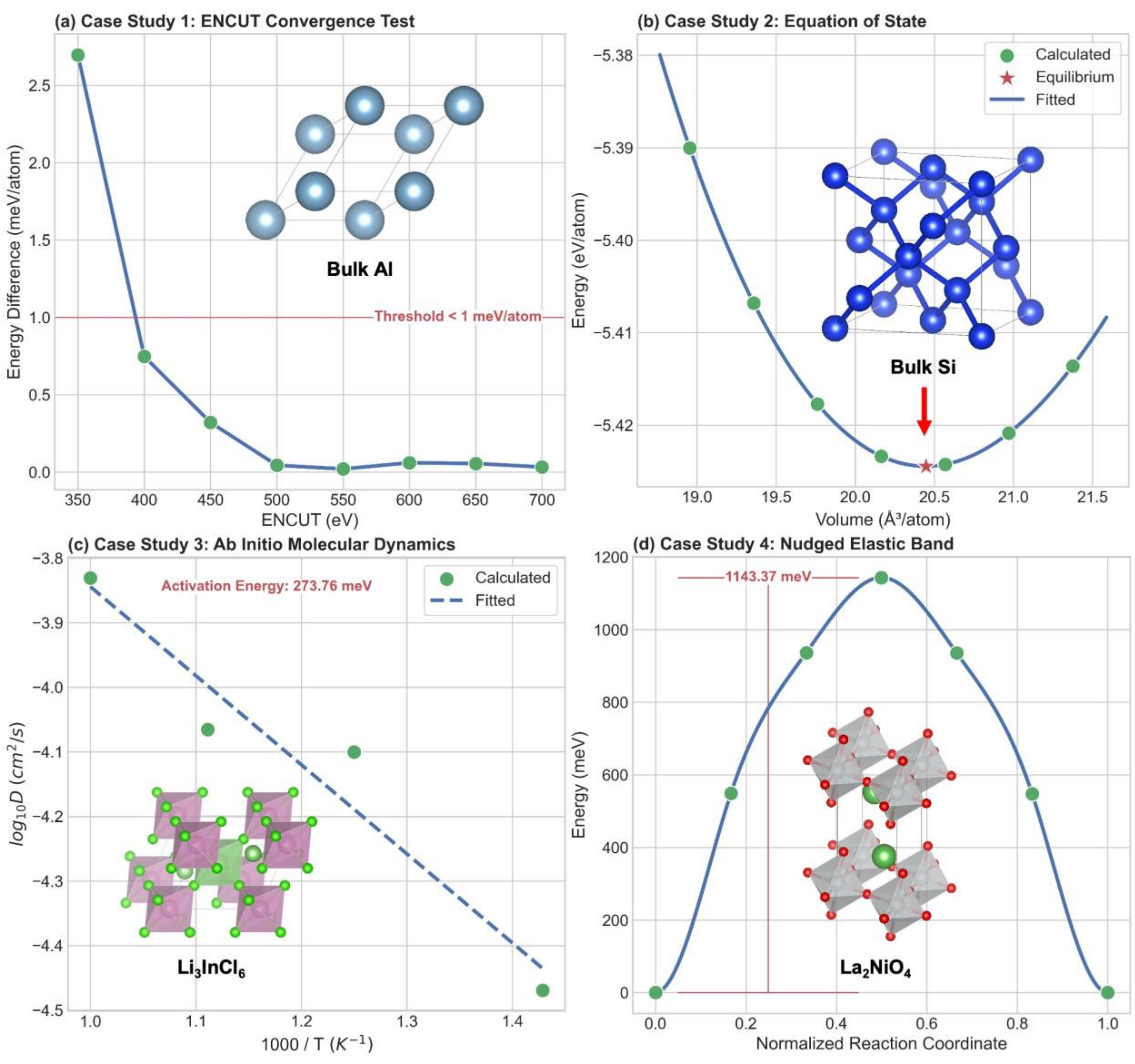

*Figure 12 Case studies demonstrating automated DFT workflows in Masgent. (a) ENCUT convergence test for bulk Al, showing the energy difference between successive plane-wave cutoffs. Convergence is reached once ΔE < 1 meV/atom (red dashed line). (b) EOS calculation for bulk Si, where DFT-calculated energies (green points) are fitted using the Birch-Murnaghan (blue curve) to obtain the equilibrium volume (red star). (c) AIMD simulation for $Li_2InCl_6$ showing the Arrhenius behavior of the Li diffusion coefficient as a function of inverse temperature. (d) NEB calculation illustrating a migration pathway between initial and final configurations. The spline-smoothed energy profile (blue curve) shows the normalized reaction coordinate and associated energy barrier.*

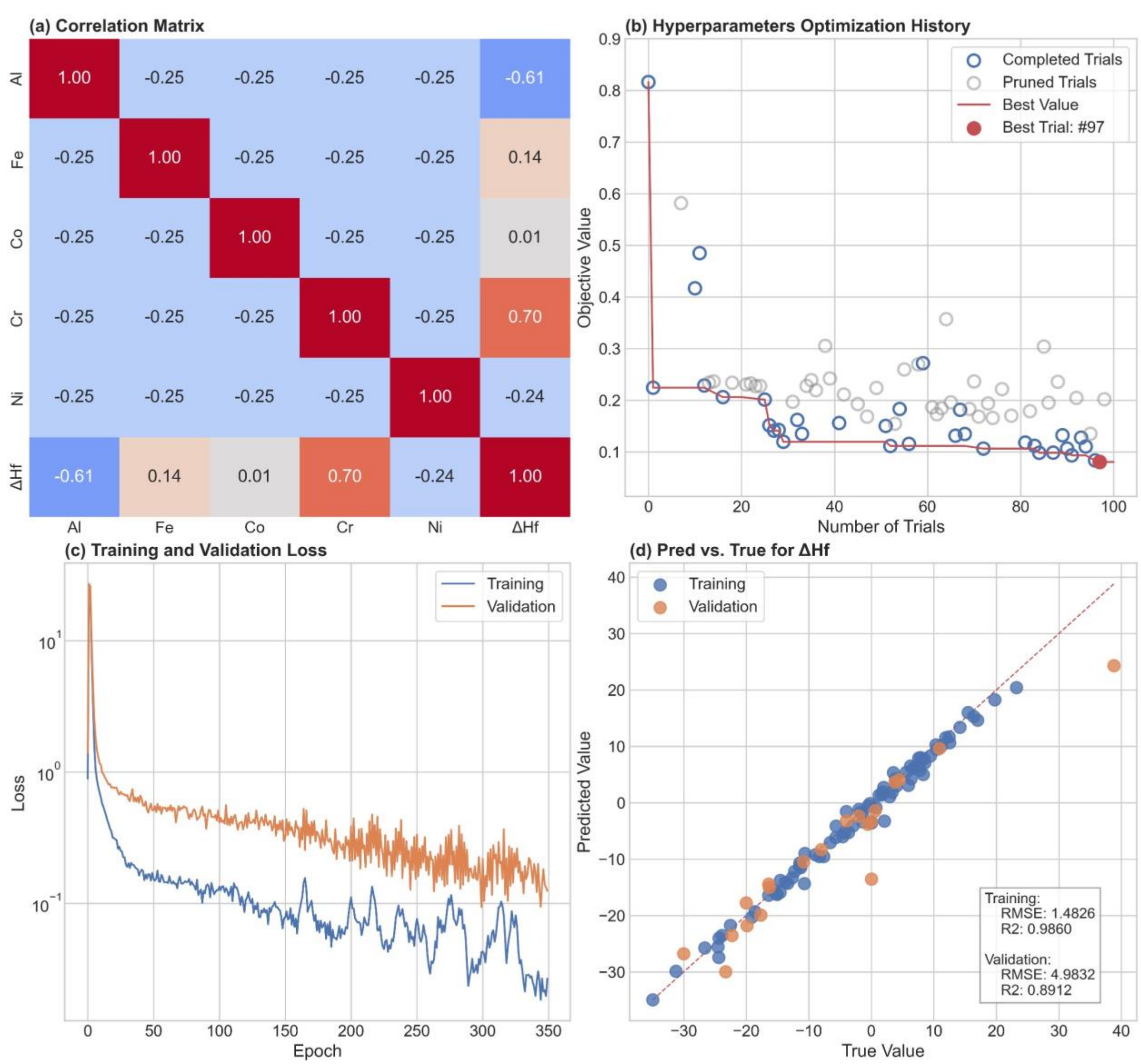

*Figure 13 Case study demonstrating ML modeling of formation enthalpies ($\Delta H_f$) for high-entropy Al-Co-Cr-Fe-Ni alloys. (a) Correlation matrix showing linear relationships among elemental fraction descriptors and the target $\Delta H_f$. (b) Optuna hyperparameter optimization history, illustrating completed and pruned trials and highlighting the best-performing configuration. (c) Training and validation loss curves with early stopping, demonstrating smooth convergence and model stabilization. (d) Predicted versus true $\Delta H_f$ for both training and validation datasets, with corresponding RMSE and $R^2$ metrics indicating strong model performance.*